\documentclass[12pt]{iopart_BW}
\usepackage{color}
\usepackage{hyperref}
\usepackage{graphicx}
\usepackage{gensymb}
\usepackage{textcomp}
\usepackage{setspace}
\usepackage[square,comma,sort&compress,numbers]{natbib}
\usepackage{fancyhdr}
\usepackage{setspace}
\usepackage{booktabs}
\usepackage{amssymb}
\usepackage{amsmath}

\newcommand{\figuremacro}[4]{
	\begin{figure}[tbp]
		\centering
		\includegraphics[width=#4\textwidth]{#1}
		\caption[#2]{#3}
		\label{fig:#1}
	\end{figure}
}

\begin{document}


\newcommand{\titletext}{Lattice dynamics of $\alpha$-cristobalite and the Boson peak in silica glass}
\title{\titletext}

\author{Bj\"orn Wehinger$^{1,2}$, Alexe\"i Bosak$^3$, Keith Refson$^{4}$\footnote{Present address: Department of Physics, Royal Holloway, University of London, Egham, Surrey TW20 0EX, United Kingdom}, Alessandro Mirone$^3$, Aleksandr Chumakov$^3$ and Michael Krisch$^3$}
\address{$^1$ Department of Quantum Matter Physics, University of Geneva, 24, Quai Ernest Ansermet, CH-1211 Gen\`eve, Switzerland}
\address{$^2$ Laboratory for Neutron Scattering and Imaging, Paul Scherrer Institute, CH-5232 Villigen PSI, Switzerland}
\address{$^3$ ESRF - The European Synchrotron, 71, Avenue des Martyrs, F-38000 Grenoble, France}
\address{$^4$ STFC Rutherford Appleton Laboratory, Harwell Science and Innovation Campus, Oxfordshire OX11 0QX, United Kingdom}

\ead{bjorn.wehinger@unige.ch}

\begin{abstract}
The lattice dynamics of the silica polymorph $\alpha$-cristobalite has been investigated by a combination of diffuse and inelastic x-ray scattering and \textit{ab initio} lattice dynamics calculations. Phonon dispersion relations and vibrational density of states are reported and the phonon eigenvectors analysed by a detailed comparison of scattering intensities. The experimentally validated calculation is used to identify the vibration contributing most to the first peak in the density of vibrational states. The comparison of its displacement pattern to the silica polymorphs $\alpha$-quartz and coesite and to vitreous silica reveals a distinct similarity and allows for decisive conclusions on the vibrations causing the so-called Boson peak in silica glass.
\end{abstract}


\maketitle

\tableofcontents

\pagestyle{fancy}
\setlength{\headheight}{16pt}
\lhead{\textit{\titletext}}
\rhead{\thepage}
\cfoot{}

\section{Introduction}
Cristobalite, SiO$_2$, crystallises at high temperature to a cubic structure ($\beta$-phase) and undergoes a phase transition to its $\alpha$-phase upon cooling. $\alpha$-cristobalite has space group P4$_1$2$_1$2 and is stable at room temperature and ambient pressure \cite{pluth_jap_1985}. Natural crystals are usually twinned \cite{dollase_zkri_1965}. The lattice dynamics of $\alpha$-cristobalite has attracted considerable interest due to its auxetic behaviour at molecular level \cite{yeganeh_science_1992, kimizuka_prl_2000, alderson_prl_2002} and the eventual realisation of a true surface wave occurring at a single isolated point within a bulk band \cite{truzupek_prl_2009}. 
It furthermore exhibits a mass density (2.289(1) g/cm$^3$) very comparable to ambient silica glass (2.20(1) g/cm$^3$) and very similar thermodynamics properties \cite{chumakov_prl_2014}. It could in fact be shown, that the excess states over the Debye law in the vibrational density of states (VDOS) in absolute numbers are very similar for $\alpha$-cristobalite and ambient silica glass. The same is true for $\alpha$-quartz and densified silica glass with matched densities. These excess states, which give rise to the so-called 'Boson peak' in glassy systems, are located at the same energies and provide the same heat capacity as in the crystals with matched densities. These observations strongly suggest that, similar to crystals, the excess of vibrational states in glasses originates from the piling up of the acoustic-like branches near the boundary of the pseudo-Brillouin zone. A detailed study of the lattice dynamics of $\alpha$-cristobalite is thus an essential prerequisite to find analogies and gain further inside into the atomic dynamics and thermodynamics of ambient silica glass.
So far, the lattice dynamics of $\alpha$-cristobalite has been investigated by inelastic neutron, Brillouin and Raman scattering. Phonon frequencies at the $\Gamma$-point, powder averaged neutron scattering spectra, elastic constants and thermodynamics properties could be determined \cite{leadbetter_jcp_1969, dove_prl_1997, bates_jcp_1972,sigaev_ncs_1999,swainson_pcm_2003,papst_cs_2013}. A measurement of the full dispersion relation and the investigation of the vibration contributing most to the first peak in the VDOS is not reported. 
Phonon dispersion relation were calculated using a Born-von-Karman force model \cite{hua_pcs_1989}, while first principles studies are limited to the $\Gamma$-point \cite{coh_prb_2008}.

In this study we aim at a detailed examination of the lattice dynamics of $\alpha$-cristobalite and a connection to the Boson peak in ambient silica glass. Dynamical matrices and derived properties are calculated from first principles and compared to experiment. The distribution of low energy features over reciprocal space is analysed by x-ray diffuse scattering. The scattering intensities are confronted to calculated thermal diffuse scattering (TDS) and compared to predictions from a rigid unit model \cite{dove_jcm_2007} and electron diffraction \cite{withers_pcm_1989}. Selected regions in reciprocal space are further investigated using inelastic x-ray scattering (IXS) which is also used to measure the dispersion relation along selected directions and the VDOS. The vibration contributing most to the first peak in the VDOS is identified with help of the validated calculation and compared to similar features in $\alpha$-quartz and coesite. Finally, the observed displacement patterns are put into relation with atomic motions reported for vitreous silica \cite{buchenau_prl_1984} to make the connection to the Boson peak.

\section{Experimental Details}
Natural $\alpha$-cristobalite crystals from the Ellora Caves, Hyderabad, India used in this study were kindly made available from the Harvard Mineralogical Museum collection. An octahedral single crystal with 0.2 mm along the four-fold axis was used for both diffuse and inelastic x-ray scattering studies, which were conducted at room temperature. One major twin and some mosaic spread were observed during the x-ray diffuse scattering. One domain could be isolated for the experimental study thanks to the small x-ray beam size, and only a small contribution from the other twins was observed. A synthetic polycrystalline sample (Materials Research Institute and School of Physics and Astronomy, Queen Mary University of London, UK) was used for the powder IXS measurements. The purity of the polycrystalline sample was verified by high-resolution x-ray diffraction on beamline ID28 at the ESRF - The European Synchrotron. The samples revealed pure single-phase patterns. The crystalline quality was checked by x-ray diffuse scattering. Clear patterns of diffraction rings without noticeable effects of structural disorder were observed.

The x-ray diffuse scattering experiment was conducted on beamline ID29 \cite{deSanctis_jsr_2012} at the ESRF. Monochromatic X-rays with 17.712 keV energy were scattered from the crystal in transmission geometry. The sample was rotated with an increment of 0.1$^{\circ}$ orthogonal to the beam direction over an angular range of 360$^{\circ}$ while diffuse scattering patterns were recorded in shutterless mode with a PILATUS 6M detector \cite{kraft_jsr_2009}. The orientation matrix and geometry of the experiment were refined using the CrysAlis software package. 2D reconstructions were prepared using locally developed software.

The single crystal IXS study was carried out at ID28. The spectrometer was operated at 17.794 keV incident energy, providing an energy resolution of 3.0 meV full-width-half-maximum and a horizontal momentum resolution of 0.3 nm$^{-1}$. Energy transfer scans were performed at constant momentum transfer (Q) in transmission geometry along selected directions in reciprocal space. Further details of the experimental setup and the data treatment can be found elsewhere \cite{kirsch_Springer_2007}.

The powder IXS study was conducted at ID28, too. IXS spectra were taken at momentum transfers chosen away from the Debye-Scherrer rings. The Q-range covered 10 to 70 nm$^{-1}$. The data combine measurements with 1.4 meV resolution at 23.725 keV incident energy within -25 to 25 meV and 0.2 meV energy steps and measurement with 3.0 meV resolution at 17.794 keV incident energy within -25 to 180 meV and 0.7 meV steps. The elastic peak in the IXS spectra was subtracted using the instrumental resolution function determined from a polymethylmethacrylate (PMMA) sample at momentum transfers close to the maximum of its structure factor. The generalised x-ray weighted VDOS (X-VDOS) was obtained by summing IXS spectra within the incoherent approximation following a previously described data treatment procedure \cite{bosak_prb_2005}. 

\section{Calculation}
The lattice dynamics calculations were performed using density functional perturbation theory (DFPT) \cite{gonze_prb_1997_2} as implemented in the CASTEP code \cite{clark_zkri_2005,refson_prb_2006}. We apply the local density approximation (LDA) with the exchange correlation functional by Perdew and Zunger \cite{perdew_prb_1981} and use the plane-wave formalism with norm-conserving pseudopotentials of the optimised form \cite{rappe_prb_1990}. Three reference orbitals were treated as valence states for Silicon and two for Oxygen. The plane wave cut-off and the sampling of the electronic grid were carefully tested by evaluating the convergence of internal forces. The electronic structure was computed on a $6 \times 6 \times 6$ Monkhorst-Pack grid and the plane wave cut-off was set to 800 eV. The chosen settings provide a convergence of internal forces to $< 10^{-3}$ eV/\AA. The cell geometry was optimised employing the Broyden-Fletcher-Goldfarb-Shannon method \cite{pfrommer_jcp_1997} by varying lattice and internal parameters. For the cell parameters of the optimised cell we find a = b = 4.939 \AA \hspace{1pt} and c = 6.870 \AA \hspace{1pt}. The obtained values compare within 1.1\% to those determined by x-ray diffraction (a = b = 4.97(8) \AA \hspace{1pt} and c = 6.94(8) \AA \hspace{1pt} \cite{dollase_zkri_1965}).
Phonon frequencies and eigenvectors were computed on a $7 \times 7 \times 7$ Monkhorst-Pack grid of the irreducible part of the Brillouin zone by perturbation calculations. Sum rules for the acoustic branches as well as the charge neutrality at the level of Born effective charges were imposed with a maximum correction of 1.2 meV at $\Gamma$. A Fourier interpolation with a grid spacing of 0.005 \AA$^{-1}$ \hspace{1pt} in the cumulant scheme including all image force constants was applied for the VDOS \cite{gonze_prb_1994,parlinski_prl_1997}. The calculation was tested to be well converged with a maximum error in phonon energies of $<$ 0.05 meV. 

Scattering intensities for inelastic and thermal diffuse scattering were calculated in first order approximation assuming the validity of both harmonic and adiabatic approximation. The formalism can be found elsewhere \cite{kirsch_Springer_2007,xu_zkri_2005}. The momentum transfer dependence of the atomic scattering factors was taken into account using an analytic function with coefficients derived from Hartree-Fock wavefunctions \cite{cromer_ac_1968}. IXS spectra are resolved in phonon energy and the intensity is sensitive to the eigenvectors. TDS intensities are not resolved in energies, but due to their strong energy dependence, they are sensitive to low energy features.

\section{Results}
High symmetry reciprocal space sections of experimental diffuse scattering and calculated TDS intensity distributions are shown in Figure \ref{fig:cristobalite_tds}. A complex distribution of diffuse scattering is noticeable. The experimental intensity distributions show signatures of the mosaic spread, which appear as arclike features and small peaks aligned on Debye-Scherrer rings. Strong diffuse scattering of acoustic phonons is noted around the Bragg peaks and streaks of diffuse intensities are visible in \mbox{$<$1 0 0$>$} and \mbox{$<$1 1 0$>$} direction in the HK0 plane. The H0L plane shows diffuse scattering along \mbox{$<$1 0 1$>$} which is more intense for low energy transfers, in particular between \mbox{($\bar{2}$ 0 0)} and \mbox{(0 0 2)}. Its origin was proven by IXS to contain a strong elastic contribution which corresponds to the inter-twin boundary and inelastic contribution. In the HHL plane we note diffuse features along \mbox{$<$1 1 0$>$} and even stronger streaks are observed along \mbox{$<$1 1 2$>$}. The diffuse scattering along \mbox{$<$1 1 0$>$} and \mbox{$<$1 0 1$>$} are in agreement with the prediction of the rigid unit modes model \cite{dove_jcm_2007}. In this model the lines of diffuse features correspond to the zero-frequency solution to the dynamical equations for an infinite framework of rigid corner-linked tetrahedra. Diffuse scattering along these directions were also observed by electron diffraction \citep{withers_pcm_1989}. The TDS intensity distribution shows features arising from low energy phonons. The slight deformation of the tetrahedra results in positive frequency solutions. The overall agreement between experiment and \textit{ab initio} calculation is good and shows that the eigenvectors of the low energy phonons are well reproduced across reciprocal space. The calculated TDS intensity distribution in 3D around \mbox{(3 3 2)} is shown in Figure \ref{fig:cristobalite_tds}. Around this Bragg peak the low energy features are particularly pronounced in all directions. Their distribution in 3D reciprocal space  partially agrees with the prediction from the rigid unit mode model but is certainly more complex and the shape differs significantly in particular along \mbox{$<$1 1 0$>$}. This may be appreciated by comparing Figure \ref{fig:cristobalite_tds} with the distribution of rigid unit modes wavevectors (Figure 12 in Reference \cite{dove_jcm_2007}). 
\figuremacro{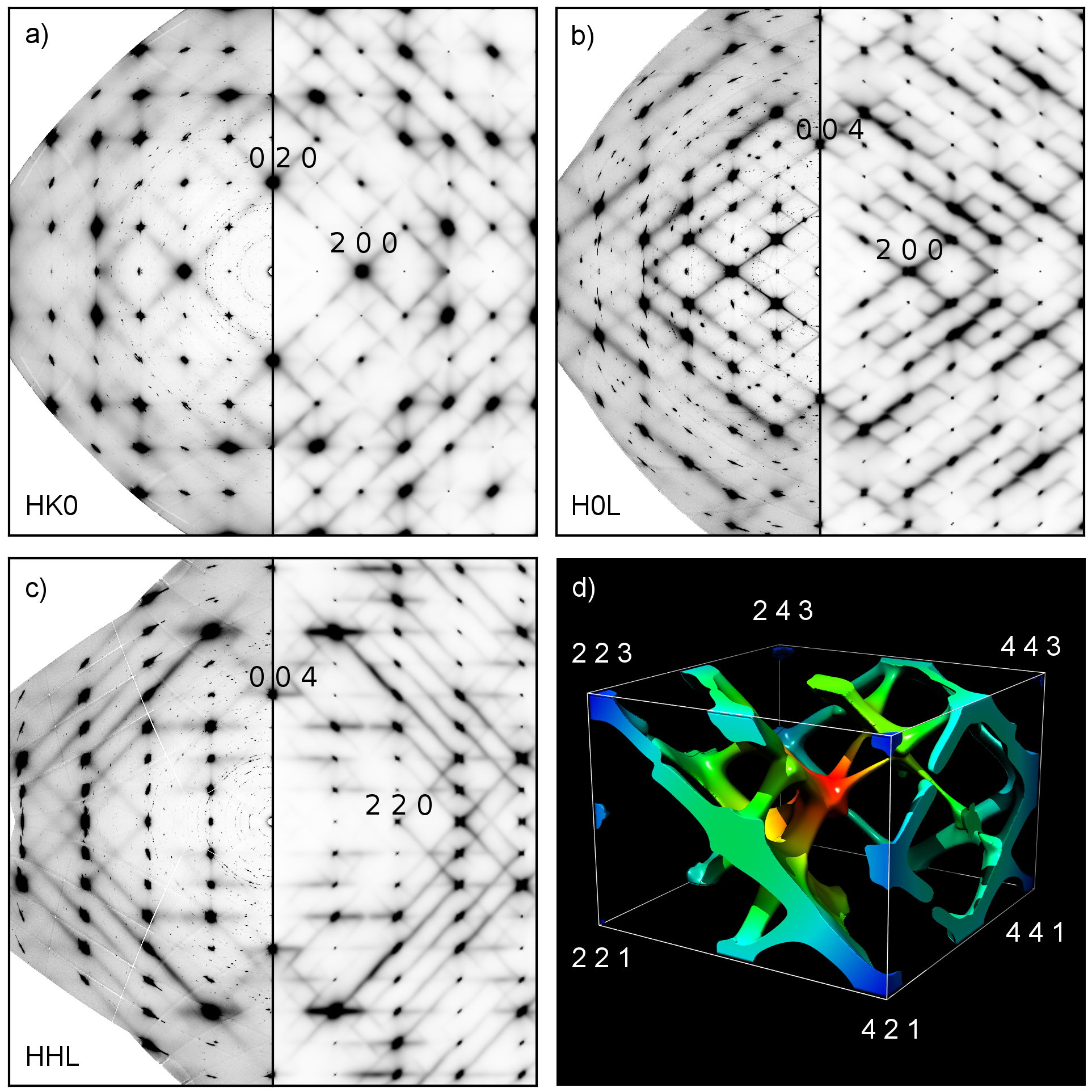}{}{(a - c) Experimental diffuse scattering (left part of individual panels) and calculated (right part of individual panels) TDS intensity distribution of single crystal $\alpha$-cristobalite in the indicated reciprocal space sections. (d) Calculated iso-intensity distribution of TDS around \mbox{(3 3 2)}. The colours denote the distance from the center. See text for further details.}{1.0}
\figuremacro{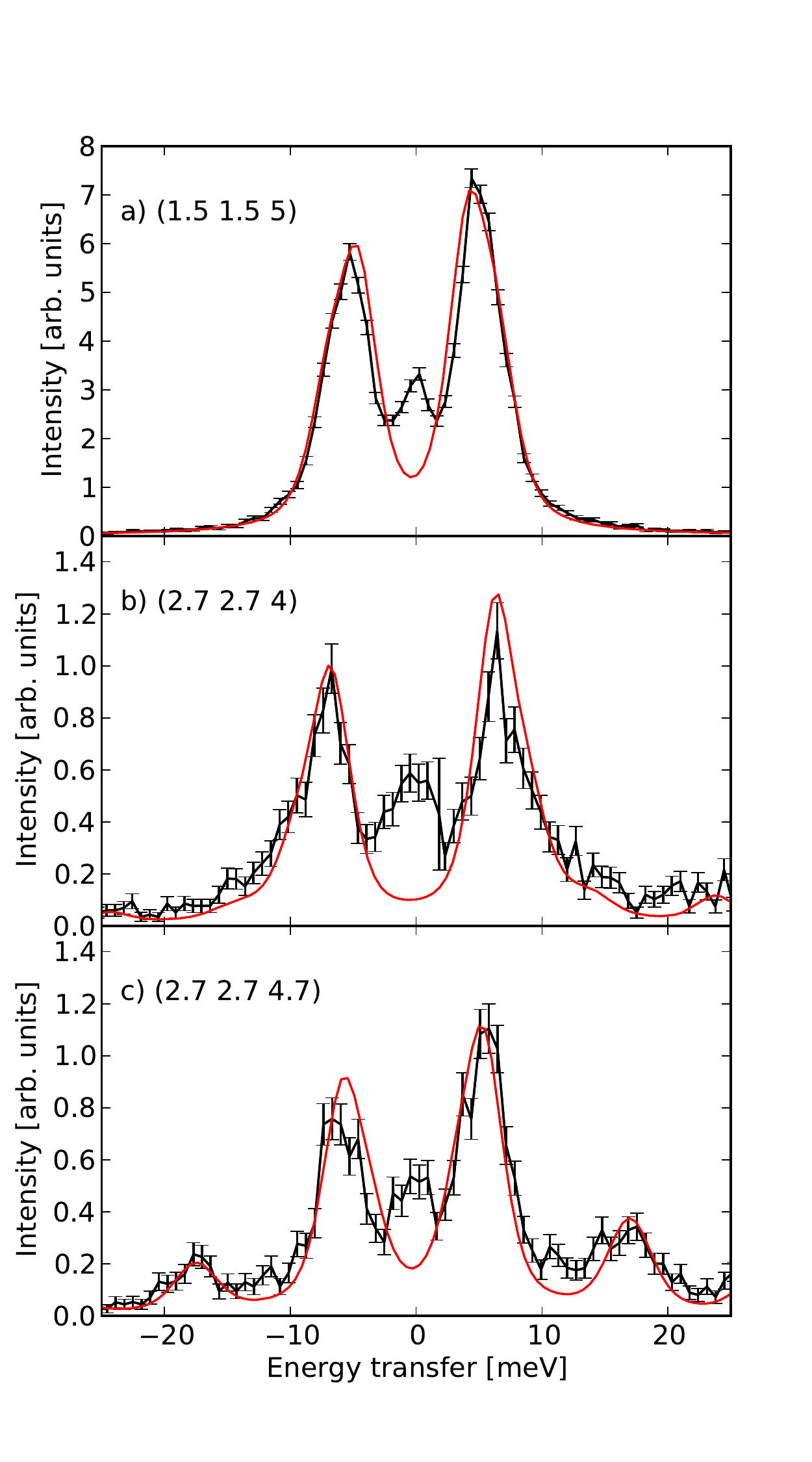}{}{Experimental (black points with error bars) and theoretical (red line) IXS spectra of single crystal $ \alpha $-cristobalite at the indicated reciprocal space points. Theoretical intensities were convoluted with the experimental resolution function and the energy transfer was scaled by 1.039. The excitation in the spectrum at the M point (a) contains the contribution of two branches, see Figure \ref{fig:cristobalite_dispersion}.}{0.6}
\figuremacro{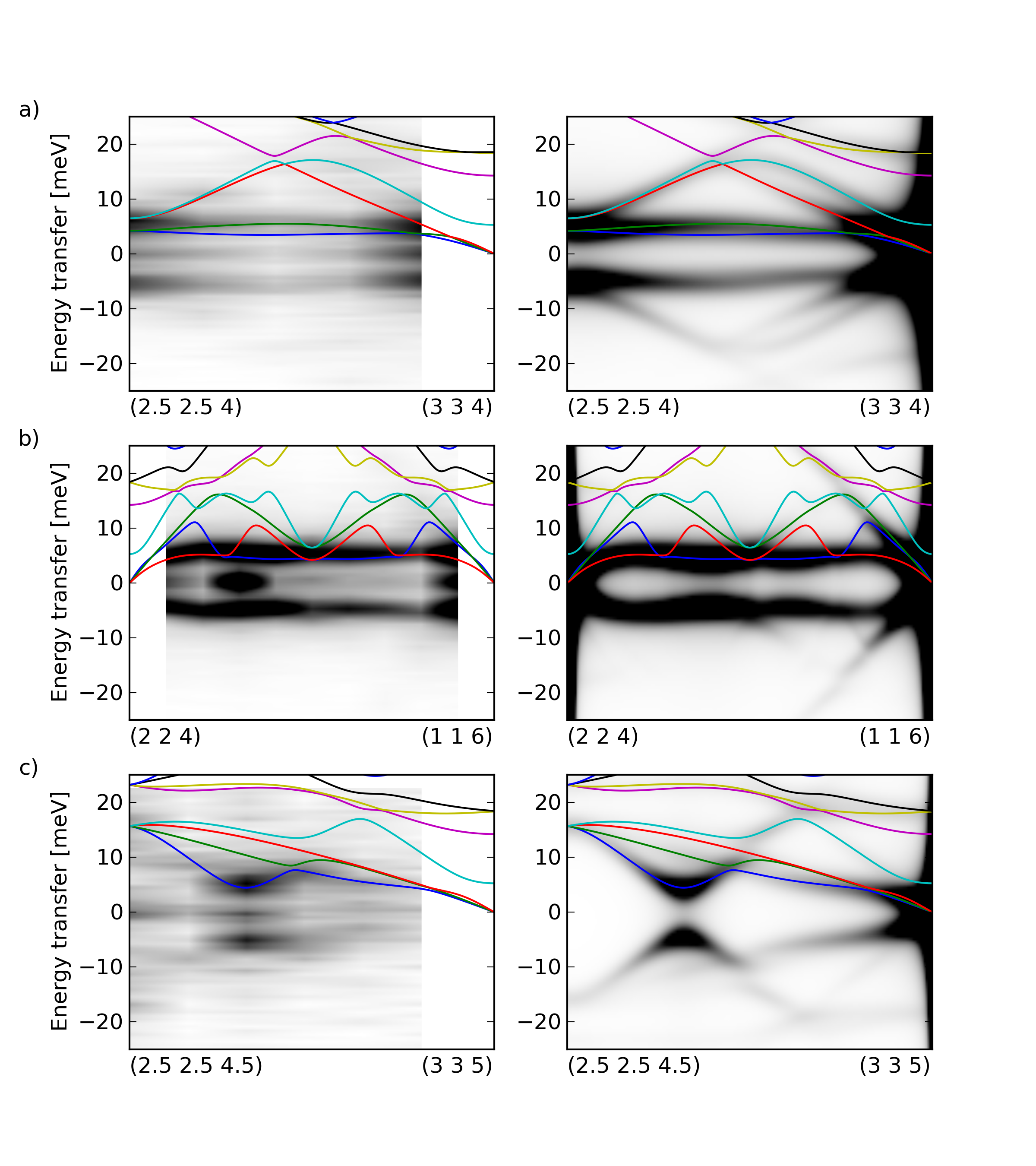}{}{Experimental IXS intensity maps (left panels) from $ \alpha $-cristobalite crystal together with theoretical intensity maps (right panels) along \mbox{$<$1 1 0$>$} (a), \mbox{$<$1 1 2$>$} (b) and \mbox{$<$1 1 1$>$} (c). The experimental maps in a) consist of 5 and in b) and c) of 8 measured spectra with linear $ \boldsymbol{q} $-spacing and energy steps of 0.7 meV. The momentum- and energy- transfers are linearly interpolated to 200 q-points and 72 energy steps. The theoretical dispersion relations are traced as lines. The theoretical maps were calculated from the eigenvectors and eigenfrequencies for 200 points along a given direction in reciprocal space and convoluted with the experimental resolution function of the spec\-tro\-meter. The absolute intensity is scaled for best visualisation of the inelastic features.}{0.8}

The streaks of diffuse scattering along \mbox{$<$1 1 0$>$} and \mbox{$<$1 1 2$>$} and some more directions were explored by IXS.
A few prototypical scans are shown in Figure \ref{fig:cristobalite_ixs} and compared to calculated spectra. Some more IXS scans are summarised in the intensity maps along certain directions in Figure \ref{fig:cristobalite_ixs_maps}. The spectra are reasonably well reproduced by the calculation. Applying a uniform scaling factor of 1.039 to the calculated phonon energies both energies and intensity of the phonon excitations match. The scaling factor was determined from the X-VDOS, and its value is justified further below. The minima of the lowest energy phonon branch are slightly underestimated in the showcases. A small elastic line is present in all spectra and not reproduced by the calculation. It arises from crystal defects. Figure \ref{fig:cristobalite_ixs_maps}a and b show the IXS intensity map along \mbox{$<$1 1 0$>$} and \mbox{$<$1 1 2$>$}, respectively. We note quite flat phonon branches with energies around 5 meV which give rise for the diffuse scattering. The calculation underestimates the energies of the lowest energy phonon bands slightly.

\figuremacro{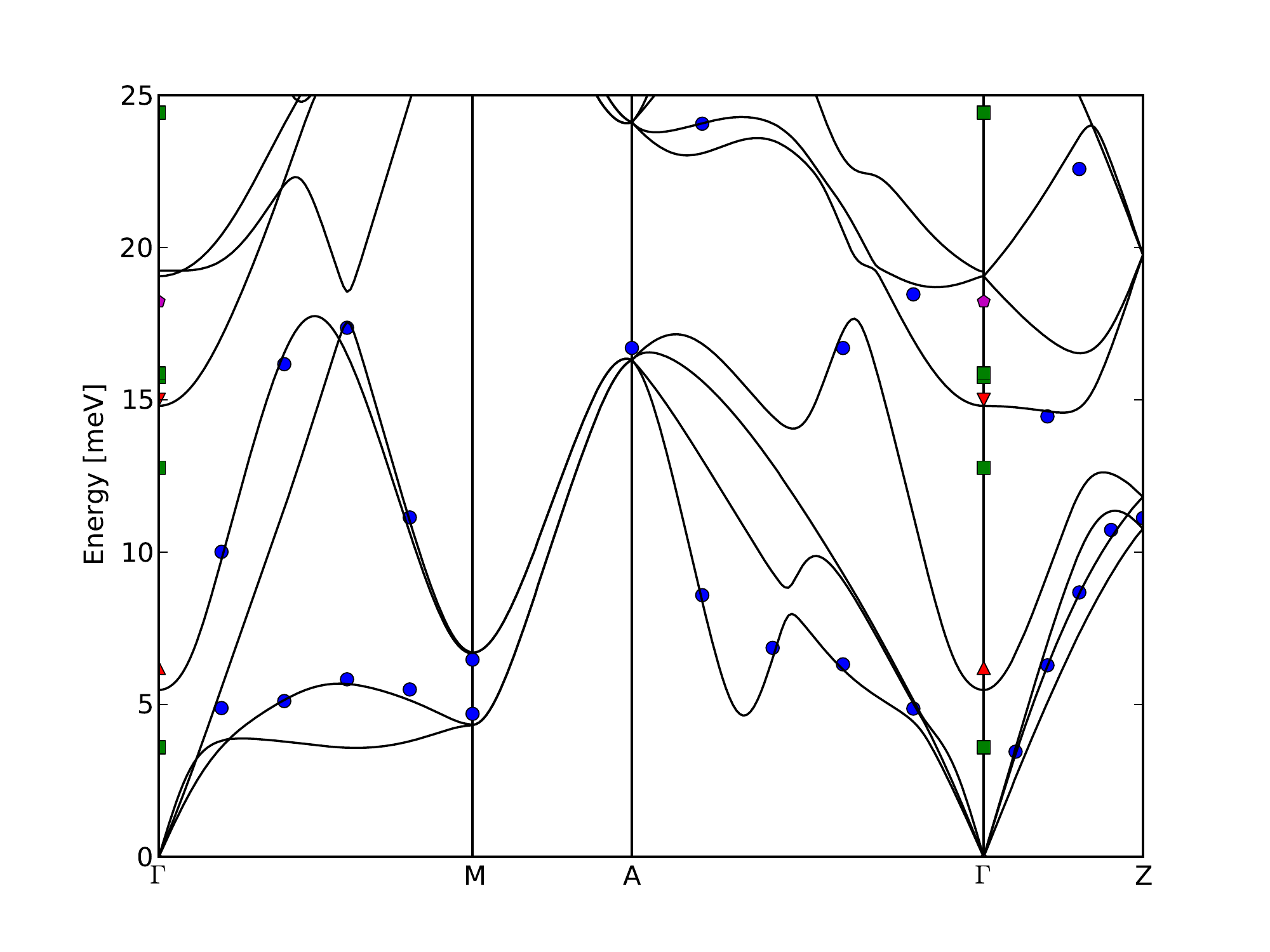}{}{Phonon dispersion relations of $\alpha $-cristobalite along high symmetry directions as obtained from the calculation (black lines) and the single crystal IXS experiment (blue points). The phonon energies at the $ \Gamma $ point are compared to Infrared \cite{swainson_pcm_2003} (magenta diamonds) and Raman measurements \cite{bates_jcp_1972} (red triangles up) and \cite{sigaev_ncs_1999} (red triangles down) as well as \textit{ab initio} calculated values from \cite{coh_prb_2008} (green squares). The energies of our calculation are scaled by 1.039.}{0.8}

Calculated dispersion relations along high symmetry directions together with experimental values from the IXS measurements are shown in Figure \ref{fig:cristobalite_dispersion}. The phonon energies at the $ \Gamma $ point are compared to infrared \cite{swainson_pcm_2003} and Raman measurements \cite{bates_jcp_1972,sigaev_ncs_1999} as well as to \textit{ab initio} calculated values of a previous work \cite{coh_prb_2008}. We note good agreement of our calculation with all experimental values. In average the calculated values are less then 0.3 meV off, in the worst case 0.9 meV. The values from the previous calculation are significantly lower. We have performed a $\Gamma$-point calculation within the general gradient approximation in Perdew-Becke-Ernzerhof parameterization as applied in the previous work rather then LDA and used the convergence criteria of our work. The resulting phonon energies are very similar to the previous result. We can thus conclude that both calculations are well converged and can attribute the discrepancy to the exchange correlation functional.

\figuremacro{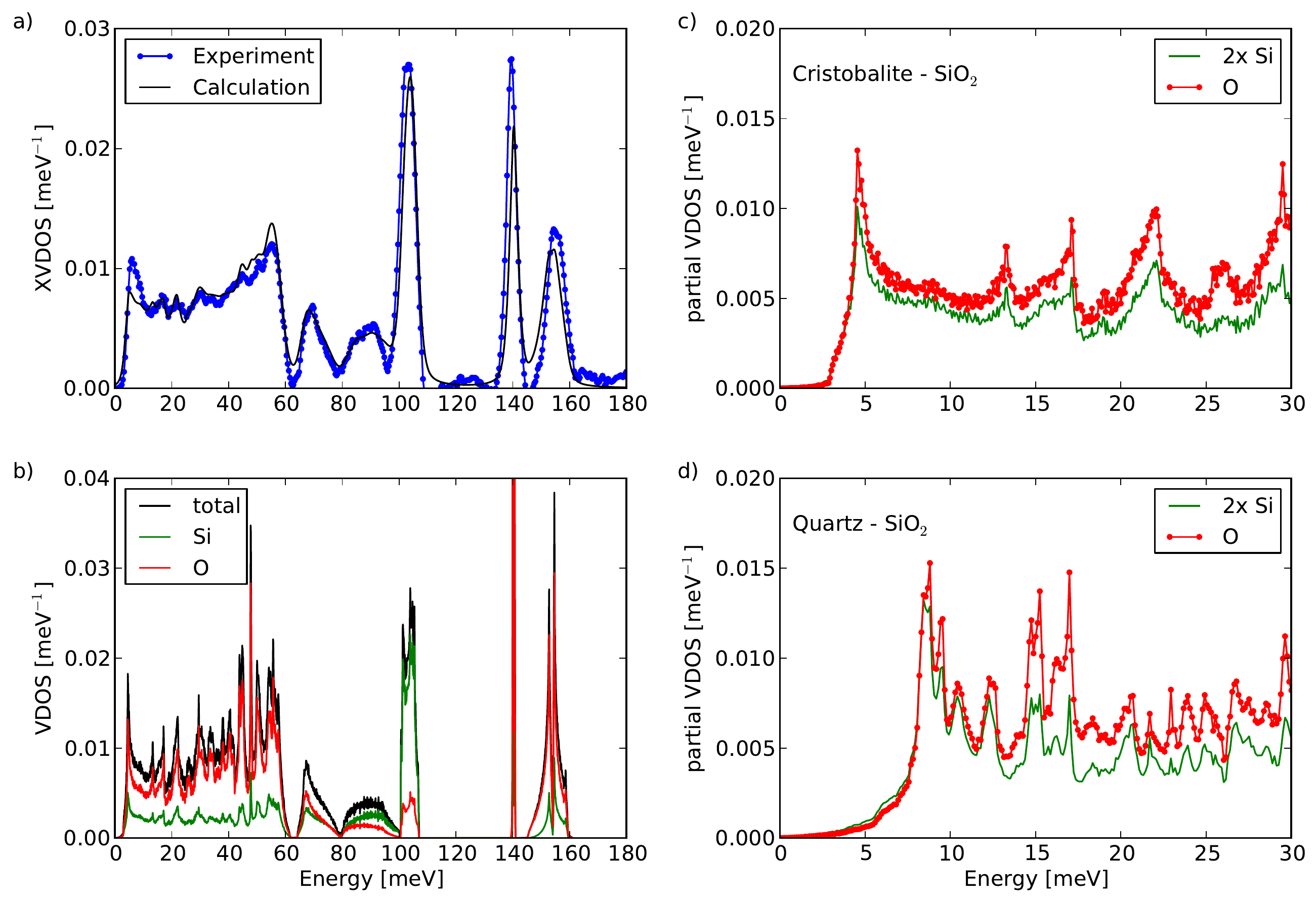}{}{a) Experimental (blue points) and calculated (black line) X-VDOS of $ \alpha $-cristobalite. The calculated X-VDOS is convoluted with the experimental resolution function and the energies are scaled by 1.039. The partial VDOS of oxygen and silicon and the total VDOS are shown in b). The partial VDOS of oxygen (red dots) and silicon (green line) of the low energy part are compared in (c), where the silicon contribution is multiplied by a factor of two. d) Partial VDOS of oxygen (red dots) and silicon (green line) in the low-energy range of $\alpha$-quartz. The silicon contribution is multiplied by a factor of two.}{1.0}

The X-VDOS for $ \alpha $-cristobalite is shown in Figure \ref{fig:cristobalite_xvdos}a and the calculated real VDOS in Figure \ref{fig:cristobalite_xvdos}b. A linear scaling of all calculated phonon energies by 1.039 leads to an almost perfect agreement of experimental and theoretical X-VDOS over the complete energy range. The applied scaling factor is very similar to what is found for the tetrahedral coordinated silica polymorphs $\alpha$-quartz, coesite and stishovite \cite{bosak_zkri_2012,wehinger_jpcm_2013,bosak_grl_2009}. 
We observe some very small discrepancies between experiment and theory, which are mainly due to the limited accuracy of sampling the reciprocal space with powder IXS spectra. The first peak in the VDOS is observed at a slightly higher energy as calculated (the discrepancy is 0.9 meV after the applied scaling).
The partial density of states (Figure \ref{fig:cristobalite_xvdos}b) separate the contribution of silicon and oxygen atoms. Focusing on the low energy part of the partial VDOS (Figure \ref{fig:cristobalite_xvdos}c) we find that the first peak located at 4.56 meV is slightly dominated by the vibration of the oxygen atoms. The low energy VDOS of $\alpha$-quartz is plotted in Figure \ref{fig:cristobalite_xvdos}d for comparison. We note that the first peak of the VDOS of $ \alpha $-cristobalite is located at a significantly lower energy than in $\alpha$-quartz. Furthermore, in $\alpha$-cristobalite the partial contribution of oxygen atoms is slightly higher than in $\alpha$-quartz.

\figuremacro{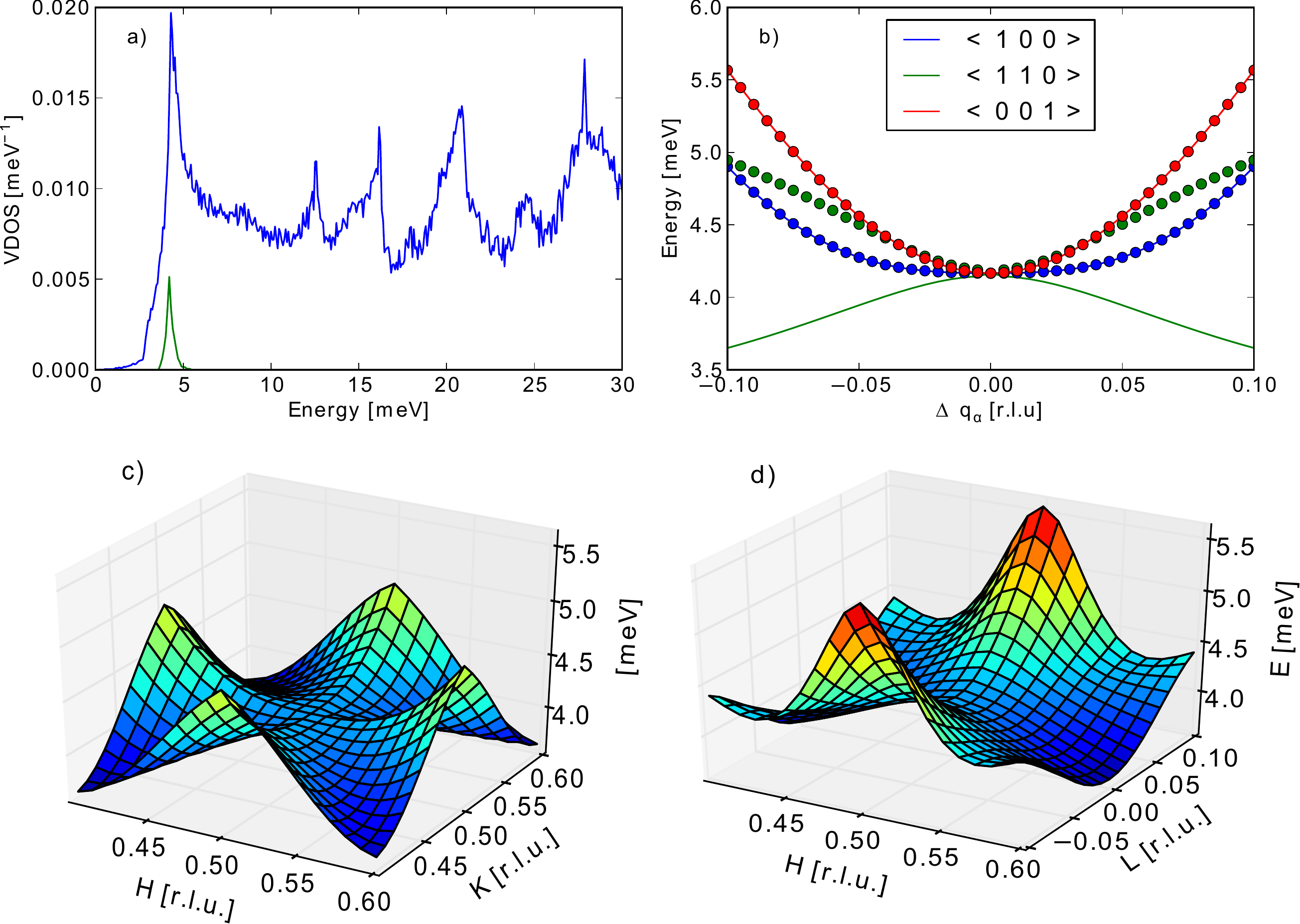}{}{Contribution and energy landscape around the critical point responsible for the first van Hove singularity in $ \alpha $-cristobalite. a) VDOS (blue) and the contribution of the energy surface containing the critical point at M (1/2 1/2 0) within $\Delta q = 1$ nm$^{-1}$ (green). b) Dispersion relations of the two branches containing the critical point along the indicated directions. The straight lines belong to the energy surface containing the saddle point; the dots belong to the sheet containing a minimum. The dotted lines are degenerate. c) and d) Energy surface projections of the energy landscape containing the saddle point in the HK0 and HHL plane, respectively.}{1.0}

The localisation of critical points contributing most to the first peak of the VDOS was determined by the simultaneous application of two filters. An energy filter of $\Delta E $= 0.3 meV was applied to the \textit{ab initio} calculated phonon energies of the first Brillouin zone, and $1/|\nabla_{\boldsymbol{q}}E(\boldsymbol{q})|$ was computed within this energy window. The investigation reveals two saddle points with the same phonon energy contributing to the first peak in the VDOS of $ \alpha $-cristobalite. The computation of the local contribution within a cube in reciprocal space of $\Delta q = 1$ nm$^{-1}$ shows that the first peak arises mainly from the region around the M point \mbox{(1/2 1/2 0)}. The local contribution is shown in Figure \ref{fig:cristobalite_cp}a. The phonon dispersion surface containing the critical point shows a double degeneracy along \mbox{$<$1 0 0$>$}, \mbox{$<$0 1 0$>$} and \mbox{$<$0 0 1$>$}, which is split along the \mbox{$<$1 1 0$>$} direction, see Figure \ref{fig:cristobalite_cp}b. The lower sheet forms a saddle point whereas the upper sheet forms a minimum which is of parabolic nature close to M. The peak in the VDOS arises thus from the lower branch for topological reasons \cite{vanhove_pr_1953}. The shape of the energy surface projected on the HK0 and HHL plane is shown in Figure \ref{fig:cristobalite_cp}c and d. A second saddle point with the same phonon energy located at \mbox{(0.47 0.2 0.33)} was found. Its contribution to the first peak in the VDOS is about a factor of two smaller compared to the contribution from the M point and less sharp.

\figuremacro{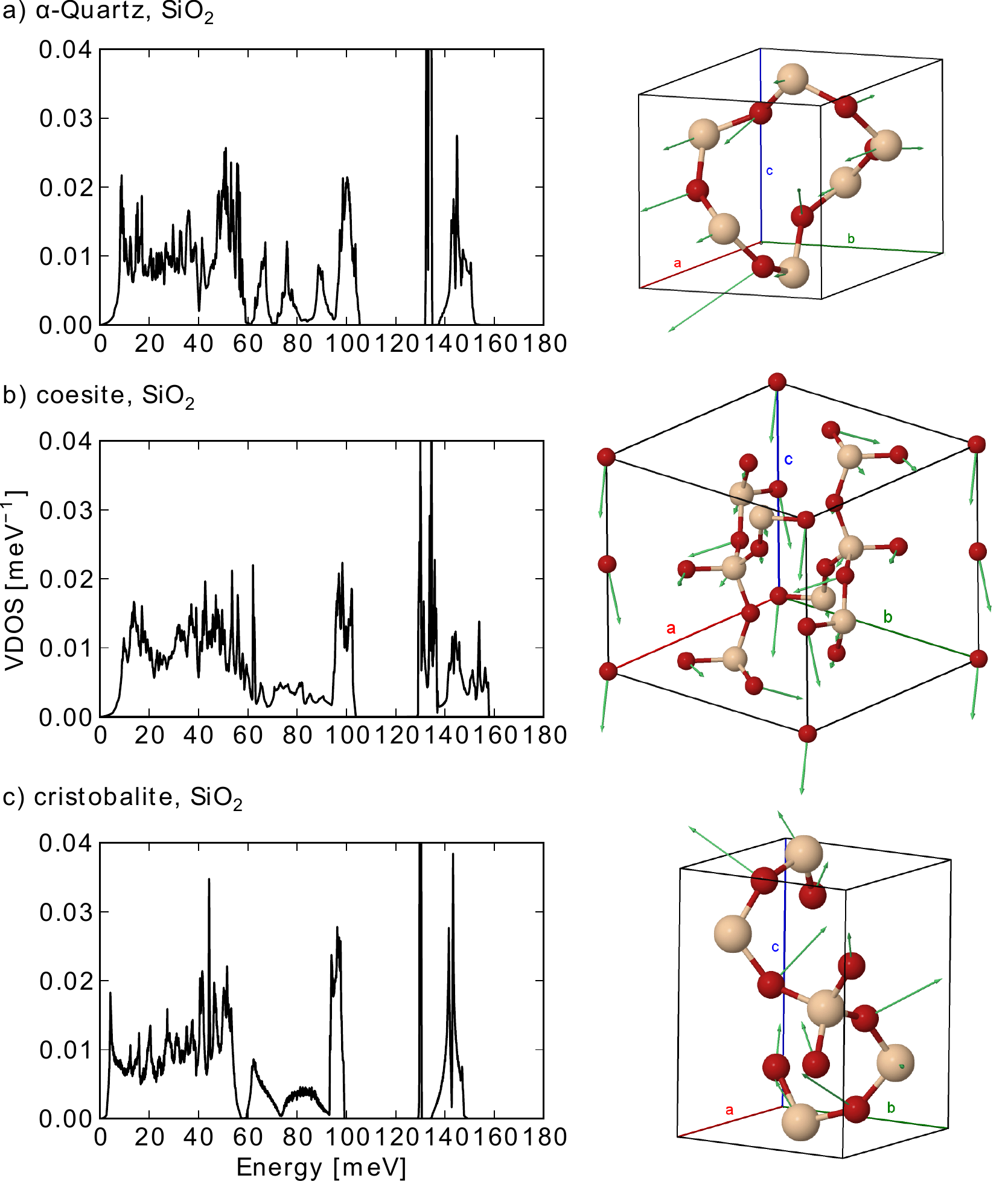}{}{Calculated VDOS of SiO$_2$ $ \alpha $-quartz (a), coesite (b) and $ \alpha $-cristobalite (c). The displacement patterns of the vibrations contributing most to the first peak in the VDOS within the unit cell are illustrated on the right (silicon atoms: beige, oxygen atoms: red). Direction and amplitude of the eigenvectors are depicted by green arrows, on arbitrary scale for better visualisation.}{1.0}

The VDOS of $ \alpha $-cristobalite and the displacement pattern contributing most to its first peak is compared in Figure \ref{fig:vdos_silica_polymorphs_atoms} with the silica polymorphs $ \alpha $-quartz and coesite \cite{wehinger_jpcm_2013}. We note, that the low energy part of the VDOS is different. In particular the first peak in the VDOS is located at different energies. The critical points which are responsible for the first peak are located at the zone boundary for all three silica polymorphs under investigation. The shape of the energy surface in the vicinity of the critical points is different, but despite this, the associated displacement patterns are very similar: The largest displacement is observed for oxygen. The vibration consists mainly of a tetrahedron tilt, accompanied by a small distortion.

\section{Discussion, Summary and Outlook}

The VDOS, obtained from powder IXS spectra, probes the ensemble of vibrational states and was used to determine an overall scaling factor for the \textit{ab initio} calculation. The application of a linear scaling factor deviating from unity by only 3.9 per cent leads to a very good agreement between \textit{ab initio} calculated and measured phonon energies. The underestimation of the calculated energies can be attributed to the limited accuracy of the exchange correlation function within the local density approximation. Detailed discussions can be found elsewhere \cite{refson_prb_2006,he_prb_2014}. 

By comparing all three, TDS intensities single crystal and powder IXS spectra over an extended region in reciprocal space, the calculation can be validated not only in terms of phonon energies but also in terms of phonon eigenvectors. The accurate determination of the complete lattice dynamics becomes possible. The full description of the lattice dynamics at arbitrary momentum transfer in turn allows for an investigation of the phonon energy landscape. The critical points responsible for Van Hove singularities can be localised by the simultaneous application of an energy and gradient filter to the calculated phonon frequencies. The eigenvectors and the shape in energy-momentum space of the vibration contributing most to the first peak in the VDOS can be assigned and compared to the tetrahedrally coordinated silica polymorphs $\alpha$-quartz and coesite. Although the first peak in the VDOS is located at different energies and the energy landscape in the vicinity of the critical points differs, a unifying picture on its origin can be drawn: The critical points are located at the zone boundary. The vibration contributing most to this first peak is mainly driven by oxygen vibrations and consists of a tetrahedron tilt, accompanied by a small distortion. Such librations of nearly undistorted SiO$_4$ tetrahedra were reported to cause the Boson peak in vitreous silica \cite{buchenau_prl_1984}. The similarity of the displacement patterns of  $\alpha$-cristobalite and vitreous silica according to Buchenaus model of coupled rotations can be appreciated in tetrahedron representation, see Figure \ref{fig:displacements_crystal_glass}. 

\figuremacro{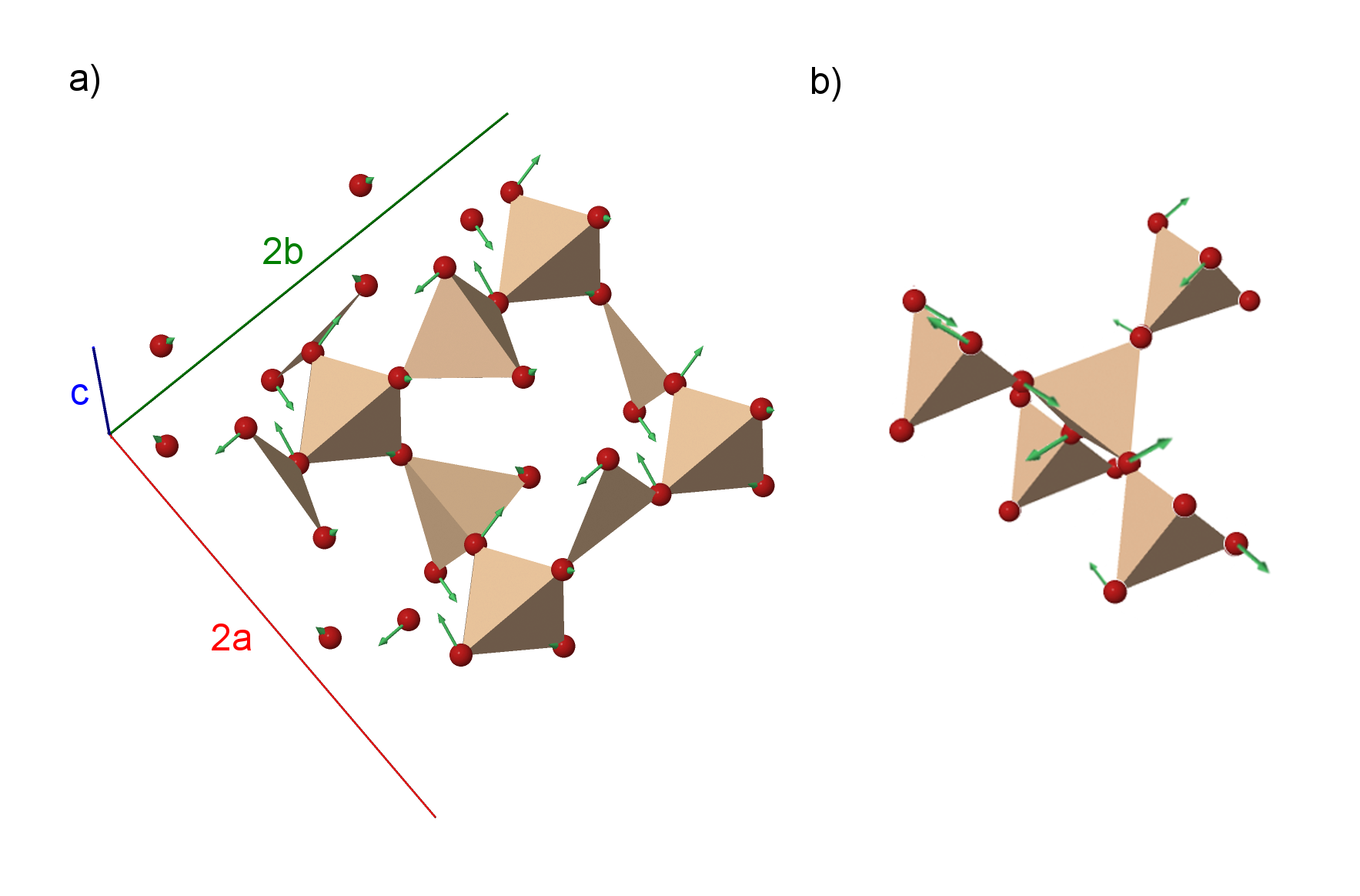}{}{Displacement patterns of SiO$_2$ thetrahedra in $\alpha$-cristobalite (a) and vitreous silica (b). $\alpha$-cristobalite is represented in a 2 $\times$ 2 $\times$ 1 supercell; the atomic motions of vitreous silica are according to the model of coupled rotations \cite{buchenau_prl_1984}.}{1.0}

Comparing the VDOS of ambient silica glass and $\alpha$-cristobalite reveals that the number of excess states above the Debye levels is very similar (6.2(9) \% and 6.8(9)\%, respectively, \cite{chumakov_prl_2014}). For both systems, the first peak in the VDOS is located nearly at the same energy. The same holds true for densified glass and $\alpha$-quartz, if the densities are matched. 
Taking into account the similarity observed in the displacement patterns as well as in the VDOS of vitreous silica and crystalline polymorphs with matched densities, we can now conclude, that the Boson peak in silica glass is the counterpart of the van Hove singularities in silica polymorphs. The vibrational states causing this feature are similar in terms of absolute numbers, energies and displacement pattern.

In summary we have seen, that the lattice dynamics of $\alpha$-cristobalite can be very well described by DFPT in LDA. The thermal diffuse scattering shows a rich structure. Some, but not all, of the strong features are in good agreement with the simplified picture of rigid unit modes. A full description of phonon frequencies and eigenvectors could be retrieved and the energy landscape of specific features in the lattice dynamics became possible. The origin of the first peak in the VDOS could be identified and linked to the Boson peak in vitreous silica.

In the framework of the silica polymorphs we are now able to provide accurate models of the lattice dynamics of $\alpha$-quartz, coesite, $\alpha$-cristobalite and stishovite \cite{bosak_zkri_2012,wehinger_jpcm_2013,bosak_grl_2009}. The calculation for these polymorphs can be extended to high pressures and allow the derivation of elastic and thermodynamical properties. Such an extension potentially builds the basis for understanding the compression mechanism and phase stability at geophysically relevant conditions. 
The prediction of the lattice dynamics should be possible within the employed calculation scheme for the high pressure phases of cristobalite \cite{dubrovinsky_cpl_2001} as well as for the recently discovered high-pressure silica polymorphs seifertite and others \cite{dubrovinsky_pepi_2004}. For cristobalite the knowledge of the lattice dynamics at high pressures is in particular important for understanding the pressure evolution of its auxetic behaviour \cite{kimizuka_mt_2005} as well as the phase stability and elasticity of its high pressure phases.

\section*{Acknowledgment}
The authors would like to acknowledge the courtesy of the Harvard Mineralogical museum and personally the collection curator Dr. Raquel Alonso-Perez for providing the high quality pieces of natural cristobalite crystals (Cristobalite, HMM\#97849) and Martin Dove (Materials Research Institute and School of Physics and Astronomy, Queen Mary University of London, UK) for the synthetic polycrystalline sample.

\def\newblock{\hskip .11em plus .33em minus .07em}
\bibliographystyle{apsrev4-1_BW}

\bibliography{references_cristobalite}

\begin{thebibliography}{40}%
\makeatletter
\providecommand \@ifxundefined [1]{%
 \@ifx{#1\undefined}
}%
\providecommand \@ifnum [1]{%
 \ifnum #1\expandafter \@firstoftwo
 \else \expandafter \@secondoftwo
 \fi
}%
\providecommand \@ifx [1]{%
 \ifx #1\expandafter \@firstoftwo
 \else \expandafter \@secondoftwo
 \fi
}%
\providecommand \natexlab [1]{#1}%
\providecommand \enquote  [1]{``#1''}%
\providecommand \bibnamefont  [1]{#1}%
\providecommand \bibfnamefont [1]{#1}%
\providecommand \citenamefont [1]{#1}%
\providecommand \href@noop [0]{\@secondoftwo}%
\providecommand \href [0]{\begingroup \@sanitize@url \@href}%
\providecommand \@href[1]{\@@startlink{#1}\@@href}%
\providecommand \@@href[1]{\endgroup#1\@@endlink}%
\providecommand \@sanitize@url [0]{\catcode `\\12\catcode `\$12\catcode
  `\&12\catcode `\#12\catcode `\^12\catcode `\_12\catcode `\%12\relax}%
\providecommand \@@startlink[1]{}%
\providecommand \@@endlink[0]{}%
\providecommand \url  [0]{\begingroup\@sanitize@url \@url }%
\providecommand \@url [1]{\endgroup\@href {#1}{\urlprefix }}%
\providecommand \urlprefix  [0]{URL }%
\providecommand \Eprint [0]{\href }%
\providecommand \doibase [0]{http://dx.doi.org/}%
\providecommand \selectlanguage [0]{\@gobble}%
\providecommand \bibinfo  [0]{\@secondoftwo}%
\providecommand \bibfield  [0]{\@secondoftwo}%
\providecommand \translation [1]{[#1]}%
\providecommand \BibitemOpen [0]{}%
\providecommand \bibitemStop [0]{}%
\providecommand \bibitemNoStop [0]{.\EOS\space}%
\providecommand \EOS [0]{\spacefactor3000\relax}%
\providecommand \BibitemShut  [1]{\csname bibitem#1\endcsname}%
\let\auto@bib@innerbib\@empty
\bibitem [{\citenamefont {Pluth}\ \emph {et~al.}(1985)\citenamefont {Pluth},
  \citenamefont {Smith},\ and\ \citenamefont {Faber}}]{pluth_jap_1985}%
  \BibitemOpen
  \bibfield  {author} {\bibinfo {author} {\bibfnamefont {J.~J.}\ \bibnamefont
  {Pluth}}, \bibinfo {author} {\bibfnamefont {J.~V.}\ \bibnamefont {Smith}}, \
  and\ \bibinfo {author} {\bibfnamefont {J.}~\bibnamefont {Faber}},\ }\bibfield
   {title} {Crystal structure of low cristobalite at 10, 293, and 473 {K}:
  {V}ariation of framework geometry with temperature,\ }\href {\doibase
  10.1063/1.334545} {\bibfield  {journal} {\bibinfo  {journal} {\emph {J. Appl.
  Phys.}}\ }\textbf {\bibinfo {volume} {57}},\ \bibinfo {pages} {1045}
  (\bibinfo {year} {1985})}\BibitemShut {NoStop}%
\bibitem [{\citenamefont {Dollase}(1965)}]{dollase_zkri_1965}%
  \BibitemOpen
  \bibfield  {author} {\bibinfo {author} {\bibfnamefont {W.~A.}\ \bibnamefont
  {Dollase}},\ }\bibfield  {title} {Reinvestigation of the structure of low
  cristobalite,\ }\href {\doibase 10.1524/zkri.1965.121.5.369} {\bibfield
  {journal} {\bibinfo  {journal} {\emph {Z. Kristallogr.}}\ }\textbf {\bibinfo
  {volume} {121}},\ \bibinfo {pages} {369} (\bibinfo {year}
  {1965})}\BibitemShut {NoStop}%
\bibitem [{\citenamefont {Yeganeh-Haeri}\ \emph {et~al.}(1992)\citenamefont
  {Yeganeh-Haeri}, \citenamefont {Weidner},\ and\ \citenamefont
  {Parise}}]{yeganeh_science_1992}%
  \BibitemOpen
  \bibfield  {author} {\bibinfo {author} {\bibfnamefont {A.}~\bibnamefont
  {Yeganeh-Haeri}}, \bibinfo {author} {\bibfnamefont {D.~J.}\ \bibnamefont
  {Weidner}}, \ and\ \bibinfo {author} {\bibfnamefont {J.~B.}\ \bibnamefont
  {Parise}},\ }\bibfield  {title} {Elasticity of $\alpha$-cristobalite: A
  silicon dioxide with a negative poisson's ratio,\ }\href {\doibase
  10.1126/science.257.5070.650} {\bibfield  {journal} {\bibinfo  {journal}
  {\emph {Science}}\ }\textbf {\bibinfo {volume} {257}},\ \bibinfo {pages}
  {650} (\bibinfo {year} {1992})}\BibitemShut {NoStop}%
\bibitem [{\citenamefont {Kimizuka}\ \emph {et~al.}(2000)\citenamefont
  {Kimizuka}, \citenamefont {Kaburaki},\ and\ \citenamefont
  {Kogure}}]{kimizuka_prl_2000}%
  \BibitemOpen
  \bibfield  {author} {\bibinfo {author} {\bibfnamefont {H.}~\bibnamefont
  {Kimizuka}}, \bibinfo {author} {\bibfnamefont {H.}~\bibnamefont {Kaburaki}},
  \ and\ \bibinfo {author} {\bibfnamefont {Y.}~\bibnamefont {Kogure}},\
  }\bibfield  {title} {Mechanism for negative poisson ratios over the $\alpha$-
  $\beta$ transition of cristobalite, {S}i{O}$_{2}$: A molecular-dynamics
  study,\ }\href {\doibase 10.1103/PhysRevLett.84.5548} {\bibfield  {journal}
  {\bibinfo  {journal} {\emph {Phys. Rev. Lett.}}\ }\textbf {\bibinfo {volume}
  {84}},\ \bibinfo {pages} {5548} (\bibinfo {year} {2000})}\BibitemShut
  {NoStop}%
\bibitem [{\citenamefont {Alderson}\ and\ \citenamefont
  {Evans}(2002)}]{alderson_prl_2002}%
  \BibitemOpen
  \bibfield  {author} {\bibinfo {author} {\bibfnamefont {A.}~\bibnamefont
  {Alderson}}\ and\ \bibinfo {author} {\bibfnamefont {K.~E.}\ \bibnamefont
  {Evans}},\ }\bibfield  {title} {Molecular origin of auxetic behavior in
  tetrahedral framework silicates,\ }\href {\doibase
  10.1103/PhysRevLett.89.225503} {\bibfield  {journal} {\bibinfo  {journal}
  {\emph {Phys. Rev. Lett.}}\ }\textbf {\bibinfo {volume} {89}},\ \bibinfo
  {pages} {225503} (\bibinfo {year} {2002})}\BibitemShut {NoStop}%
\bibitem [{\citenamefont {Trzupek}\ and\ \citenamefont
  {Zieli\ifmmode~\acute{n}\else \'{n}\fi{}ski}(2009)}]{truzupek_prl_2009}%
  \BibitemOpen
  \bibfield  {author} {\bibinfo {author} {\bibfnamefont {D.}~\bibnamefont
  {Trzupek}}\ and\ \bibinfo {author} {\bibfnamefont {P.}~\bibnamefont
  {Zieli\ifmmode~\acute{n}\else \'{n}\fi{}ski}},\ }\bibfield  {title} {Isolated
  true surface wave in a radiative band on a surface of a stressed auxetic,\
  }\href {\doibase 10.1103/PhysRevLett.103.075504} {\bibfield  {journal}
  {\bibinfo  {journal} {\emph {Phys. Rev. Lett.}}\ }\textbf {\bibinfo {volume}
  {103}},\ \bibinfo {pages} {075504} (\bibinfo {year} {2009})}\BibitemShut
  {NoStop}%
\bibitem [{\citenamefont {Chumakov}\ \emph {et~al.}(2014)\citenamefont
  {Chumakov}, \citenamefont {Monaco}, \citenamefont {Fontana}, \citenamefont
  {Bosak}, \citenamefont {Hermann}, \citenamefont {Bessas}, \citenamefont
  {Wehinger}, \citenamefont {Crichton}, \citenamefont {Krisch}, \citenamefont
  {R\"uffer}, \citenamefont {Baldi}, \citenamefont {Carini~Jr.}, \citenamefont
  {Carini}, \citenamefont {D'Angelo}, \citenamefont {Gilioli}, \citenamefont
  {Tripodo}, \citenamefont {Zanatta}, \citenamefont {Winkler}, \citenamefont
  {Milman}, \citenamefont {Refson}, \citenamefont {Dove}, \citenamefont
  {Dubrovinskaia}, \citenamefont {Dubrovinsky}, \citenamefont {Keding},\ and\
  \citenamefont {Yue}}]{chumakov_prl_2014}%
  \BibitemOpen
  \bibfield  {author} {\bibinfo {author} {\bibfnamefont {A.~I.}\ \bibnamefont
  {Chumakov}}, \bibinfo {author} {\bibfnamefont {G.}~\bibnamefont {Monaco}},
  \bibinfo {author} {\bibfnamefont {A.}~\bibnamefont {Fontana}}, \bibinfo
  {author} {\bibfnamefont {A.}~\bibnamefont {Bosak}}, \bibinfo {author}
  {\bibfnamefont {R.~P.}\ \bibnamefont {Hermann}}, \bibinfo {author}
  {\bibfnamefont {D.}~\bibnamefont {Bessas}}, \bibinfo {author} {\bibfnamefont
  {B.}~\bibnamefont {Wehinger}}, \bibinfo {author} {\bibfnamefont {W.~A.}\
  \bibnamefont {Crichton}}, \bibinfo {author} {\bibfnamefont {M.}~\bibnamefont
  {Krisch}}, \bibinfo {author} {\bibfnamefont {R.}~\bibnamefont {R\"uffer}},
  \bibinfo {author} {\bibfnamefont {G.}~\bibnamefont {Baldi}}, \bibinfo
  {author} {\bibfnamefont {G.}~\bibnamefont {Carini~Jr.}}, \bibinfo {author}
  {\bibfnamefont {G.}~\bibnamefont {Carini}}, \bibinfo {author} {\bibfnamefont
  {G.}~\bibnamefont {D'Angelo}}, \bibinfo {author} {\bibfnamefont
  {E.}~\bibnamefont {Gilioli}}, \bibinfo {author} {\bibfnamefont
  {G.}~\bibnamefont {Tripodo}}, \bibinfo {author} {\bibfnamefont
  {M.}~\bibnamefont {Zanatta}}, \bibinfo {author} {\bibfnamefont
  {B.}~\bibnamefont {Winkler}}, \bibinfo {author} {\bibfnamefont
  {V.}~\bibnamefont {Milman}}, \bibinfo {author} {\bibfnamefont
  {K.}~\bibnamefont {Refson}}, \bibinfo {author} {\bibfnamefont {M.~T.}\
  \bibnamefont {Dove}}, \bibinfo {author} {\bibfnamefont {N.}~\bibnamefont
  {Dubrovinskaia}}, \bibinfo {author} {\bibfnamefont {L.}~\bibnamefont
  {Dubrovinsky}}, \bibinfo {author} {\bibfnamefont {R.}~\bibnamefont {Keding}},
  \ and\ \bibinfo {author} {\bibfnamefont {Y.~Z.}\ \bibnamefont {Yue}},\
  }\bibfield  {title} {Role of disorder in the thermodynamics and atomic
  dynamics of glasses,\ }\href {\doibase 10.1103/PhysRevLett.112.025502}
  {\bibfield  {journal} {\bibinfo  {journal} {\emph {Phys. Rev. Lett.}}\
  }\textbf {\bibinfo {volume} {112}},\ \bibinfo {pages} {025502} (\bibinfo
  {year} {2014})}\BibitemShut {NoStop}%
\bibitem [{\citenamefont {Leadbetter}(1969)}]{leadbetter_jcp_1969}%
  \BibitemOpen
  \bibfield  {author} {\bibinfo {author} {\bibfnamefont {A.~J.}\ \bibnamefont
  {Leadbetter}},\ }\bibfield  {title} {Inelastic cold neutron scattering from
  different forms of silica,\ }\href {\doibase 10.1063/1.1672068} {\bibfield
  {journal} {\bibinfo  {journal} {\emph {J. Chem. Phys.}}\ }\textbf {\bibinfo
  {volume} {51}},\ \bibinfo {pages} {779} (\bibinfo {year} {1969})}\BibitemShut
  {NoStop}%
\bibitem [{\citenamefont {Dove}\ \emph {et~al.}(1997)\citenamefont {Dove},
  \citenamefont {Harris}, \citenamefont {Hannon}, \citenamefont {Parker},
  \citenamefont {Swainson},\ and\ \citenamefont {Gambhir}}]{dove_prl_1997}%
  \BibitemOpen
  \bibfield  {author} {\bibinfo {author} {\bibfnamefont {M.~T.}\ \bibnamefont
  {Dove}}, \bibinfo {author} {\bibfnamefont {M.~J.}\ \bibnamefont {Harris}},
  \bibinfo {author} {\bibfnamefont {A.~C.}\ \bibnamefont {Hannon}}, \bibinfo
  {author} {\bibfnamefont {J.~M.}\ \bibnamefont {Parker}}, \bibinfo {author}
  {\bibfnamefont {I.~P.}\ \bibnamefont {Swainson}}, \ and\ \bibinfo {author}
  {\bibfnamefont {M.}~\bibnamefont {Gambhir}},\ }\bibfield  {title} {Floppy
  modes in crystalline and amorphous silicates,\ }\href {\doibase
  10.1103/PhysRevLett.78.1070} {\bibfield  {journal} {\bibinfo  {journal}
  {\emph {Phys. Rev. Lett.}}\ }\textbf {\bibinfo {volume} {78}},\ \bibinfo
  {pages} {1070} (\bibinfo {year} {1997})}\BibitemShut {NoStop}%
\bibitem [{\citenamefont {Bates}(1972)}]{bates_jcp_1972}%
  \BibitemOpen
  \bibfield  {author} {\bibinfo {author} {\bibfnamefont {J.}~\bibnamefont
  {Bates}},\ }\bibfield  {title} {{R}aman spectra of $ \alpha $ and $\beta $
  cristobalite,\ }\href {\doibase 10.1063/1.1678878} {\bibfield  {journal}
  {\bibinfo  {journal} {\emph {J. Chem. Phys.}}\ }\textbf {\bibinfo {volume}
  {57}},\ \bibinfo {pages} {4042} (\bibinfo {year} {1972})}\BibitemShut
  {NoStop}%
\bibitem [{\citenamefont {Sigaev}\ \emph {et~al.}(1999)\citenamefont {Sigaev},
  \citenamefont {Smelyanskaya}, \citenamefont {Plotnichenko}, \citenamefont
  {Koltashev}, \citenamefont {Volkov},\ and\ \citenamefont
  {Pernice}}]{sigaev_ncs_1999}%
  \BibitemOpen
  \bibfield  {author} {\bibinfo {author} {\bibfnamefont {V.~N.}\ \bibnamefont
  {Sigaev}}, \bibinfo {author} {\bibfnamefont {E.~N.}\ \bibnamefont
  {Smelyanskaya}}, \bibinfo {author} {\bibfnamefont {V.~G.}\ \bibnamefont
  {Plotnichenko}}, \bibinfo {author} {\bibfnamefont {V.~V.}\ \bibnamefont
  {Koltashev}}, \bibinfo {author} {\bibfnamefont {A.~A.}\ \bibnamefont
  {Volkov}}, \ and\ \bibinfo {author} {\bibfnamefont {P.}~\bibnamefont
  {Pernice}},\ }\bibfield  {title} {Low-frequency band at 50 cm$^{-1}$ in the
  {R}aman spectrum of cristobalite: identification of similar structural motifs
  in glasses and crystals of similar composition,\ }\href {\doibase
  10.1016/S0022-3093(99)00242-2} {\bibfield  {journal} {\bibinfo  {journal}
  {\emph {J. Non-Cryst. Solids}}\ }\textbf {\bibinfo {volume} {248}},\ \bibinfo
  {pages} {141 } (\bibinfo {year} {1999})}\BibitemShut {NoStop}%
\bibitem [{\citenamefont {Swainson}\ \emph {et~al.}(2003)\citenamefont
  {Swainson}, \citenamefont {Dove},\ and\ \citenamefont
  {Palmer}}]{swainson_pcm_2003}%
  \BibitemOpen
  \bibfield  {author} {\bibinfo {author} {\bibfnamefont {I.~P.}\ \bibnamefont
  {Swainson}}, \bibinfo {author} {\bibfnamefont {M.~T.}\ \bibnamefont {Dove}},
  \ and\ \bibinfo {author} {\bibfnamefont {D.~C.}\ \bibnamefont {Palmer}},\
  }\bibfield  {title} {Infrared and {R}aman spectroscopy studies of the $
  \alpha - \beta $ phase transition in cristobalite,\ }\href {\doibase
  10.1007/s00269-003-0320-8} {\bibfield  {journal} {\bibinfo  {journal} {\emph
  {Phys. Chem. Miner.}}\ }\textbf {\bibinfo {volume} {30}},\ \bibinfo {pages}
  {353} (\bibinfo {year} {2003})}\BibitemShut {NoStop}%
\bibitem [{\citenamefont {Pabst}\ and\ \citenamefont
  {Gregorova}(2013)}]{papst_cs_2013}%
  \BibitemOpen
  \bibfield  {author} {\bibinfo {author} {\bibfnamefont {W.}~\bibnamefont
  {Pabst}}\ and\ \bibinfo {author} {\bibfnamefont {E.}~\bibnamefont
  {Gregorova}},\ }\bibfield  {title} {Elastic properties of silica polymorphs -
  a review,\ }\href@noop {} {\bibfield  {journal} {\bibinfo  {journal} {\emph
  {Ceramics-Silikaty}}\ }\textbf {\bibinfo {volume} {57}},\ \bibinfo {pages}
  {167} (\bibinfo {year} {2013})}\BibitemShut {NoStop}%
\bibitem [{\citenamefont {Hua}\ \emph {et~al.}(1989)\citenamefont {Hua},
  \citenamefont {Welberry},\ and\ \citenamefont {Withers}}]{hua_pcs_1989}%
  \BibitemOpen
  \bibfield  {author} {\bibinfo {author} {\bibfnamefont {G.}~\bibnamefont
  {Hua}}, \bibinfo {author} {\bibfnamefont {T.}~\bibnamefont {Welberry}}, \
  and\ \bibinfo {author} {\bibfnamefont {R.}~\bibnamefont {Withers}},\
  }\bibfield  {title} {Lattice dynamics of $\alpha$- and $\beta$-cristobalite,
  {S}i{O}$_2$,\ }\href {\doibase 10.1016/0022-3697(89)90419-8} {\bibfield
  {journal} {\bibinfo  {journal} {\emph {J. Phys. Chem. Solids}}\ }\textbf
  {\bibinfo {volume} {50}},\ \bibinfo {pages} {207 } (\bibinfo {year}
  {1989})}\BibitemShut {NoStop}%
\bibitem [{\citenamefont {Coh}\ and\ \citenamefont
  {Vanderbilt}(2008)}]{coh_prb_2008}%
  \BibitemOpen
  \bibfield  {author} {\bibinfo {author} {\bibfnamefont {S.}~\bibnamefont
  {Coh}}\ and\ \bibinfo {author} {\bibfnamefont {D.}~\bibnamefont
  {Vanderbilt}},\ }\bibfield  {title} {Structural stability and lattice
  dynamics of {S}i{O}$_2$ cristobalite,\ }\href {\doibase
  10.1103/PhysRevB.78.054117} {\bibfield  {journal} {\bibinfo  {journal} {\emph
  {Phys. Rev. B}}\ }\textbf {\bibinfo {volume} {78}},\ \bibinfo {pages}
  {054117} (\bibinfo {year} {2008})}\BibitemShut {NoStop}%
\bibitem [{\citenamefont {Dove}\ \emph {et~al.}(2007)\citenamefont {Dove},
  \citenamefont {Pryde}, \citenamefont {Heine},\ and\ \citenamefont
  {Hammonds}}]{dove_jcm_2007}%
  \BibitemOpen
  \bibfield  {author} {\bibinfo {author} {\bibfnamefont {M.~T.}\ \bibnamefont
  {Dove}}, \bibinfo {author} {\bibfnamefont {A.~K.~A.}\ \bibnamefont {Pryde}},
  \bibinfo {author} {\bibfnamefont {V.}~\bibnamefont {Heine}}, \ and\ \bibinfo
  {author} {\bibfnamefont {K.~D.}\ \bibnamefont {Hammonds}},\ }\bibfield
  {title} {Exotic distributions of rigid unit modes in the reciprocal spaces of
  framework aluminosilicates,\ }\href {\doibase 10.1088/0953-8984/19/27/275209}
  {\bibfield  {journal} {\bibinfo  {journal} {\emph {J. Phys.: Condens.
  Matter}}\ }\textbf {\bibinfo {volume} {19}},\ \bibinfo {pages} {275209}
  (\bibinfo {year} {2007})}\BibitemShut {NoStop}%
\bibitem [{\citenamefont {Withers}\ \emph {et~al.}(1989)\citenamefont
  {Withers}, \citenamefont {Thompson},\ and\ \citenamefont
  {Welberry}}]{withers_pcm_1989}%
  \BibitemOpen
  \bibfield  {author} {\bibinfo {author} {\bibfnamefont {R.~L.}\ \bibnamefont
  {Withers}}, \bibinfo {author} {\bibfnamefont {J.~G.}\ \bibnamefont
  {Thompson}}, \ and\ \bibinfo {author} {\bibfnamefont {T.~R.}\ \bibnamefont
  {Welberry}},\ }\bibfield  {title} {The structure and microstructure of $
  \alpha $-cristobalite and its relationship to $ \beta $-cristobalite,\ }\href
  {\doibase 10.1007/BF00202206} {\bibfield  {journal} {\bibinfo  {journal}
  {\emph {Phys. Chem. Minerals}}\ }\textbf {\bibinfo {volume} {16}},\ \bibinfo
  {pages} {517} (\bibinfo {year} {1989})}\BibitemShut {NoStop}%
\bibitem [{\citenamefont {Buchenau}\ \emph {et~al.}(1984)\citenamefont
  {Buchenau}, \citenamefont {N\"ucker},\ and\ \citenamefont
  {Dianoux}}]{buchenau_prl_1984}%
  \BibitemOpen
  \bibfield  {author} {\bibinfo {author} {\bibfnamefont {U.}~\bibnamefont
  {Buchenau}}, \bibinfo {author} {\bibfnamefont {N.}~\bibnamefont {N\"ucker}},
  \ and\ \bibinfo {author} {\bibfnamefont {A.~J.}\ \bibnamefont {Dianoux}},\
  }\bibfield  {title} {Neutron scattering study of the low-frequency vibrations
  in vitreous silica,\ }\href {\doibase 10.1103/PhysRevLett.53.2316} {\bibfield
   {journal} {\bibinfo  {journal} {\emph {Phys. Rev. Lett.}}\ }\textbf
  {\bibinfo {volume} {53}},\ \bibinfo {pages} {2316} (\bibinfo {year}
  {1984})}\BibitemShut {NoStop}%
\bibitem [{\citenamefont {de~Sanctis}\ \emph {et~al.}(2012)\citenamefont
  {de~Sanctis}, \citenamefont {Beteva}, \citenamefont {Caserotto},
  \citenamefont {Dobias}, \citenamefont {Gabadinho}, \citenamefont {Giraud},
  \citenamefont {Gobbo}, \citenamefont {Guijarro}, \citenamefont {Lentini},
  \citenamefont {Lavault}, \citenamefont {Mairs}, \citenamefont {McSweeney},
  \citenamefont {Petitdemange}, \citenamefont {Rey-Bakaikoa}, \citenamefont
  {Surr}, \citenamefont {Theveneau}, \citenamefont {Leonard},\ and\
  \citenamefont {Mueller-Dieckmann}}]{deSanctis_jsr_2012}%
  \BibitemOpen
  \bibfield  {author} {\bibinfo {author} {\bibfnamefont {D.}~\bibnamefont
  {de~Sanctis}}, \bibinfo {author} {\bibfnamefont {A.}~\bibnamefont {Beteva}},
  \bibinfo {author} {\bibfnamefont {H.}~\bibnamefont {Caserotto}}, \bibinfo
  {author} {\bibfnamefont {F.}~\bibnamefont {Dobias}}, \bibinfo {author}
  {\bibfnamefont {J.}~\bibnamefont {Gabadinho}}, \bibinfo {author}
  {\bibfnamefont {T.}~\bibnamefont {Giraud}}, \bibinfo {author} {\bibfnamefont
  {A.}~\bibnamefont {Gobbo}}, \bibinfo {author} {\bibfnamefont
  {M.}~\bibnamefont {Guijarro}}, \bibinfo {author} {\bibfnamefont
  {M.}~\bibnamefont {Lentini}}, \bibinfo {author} {\bibfnamefont
  {B.}~\bibnamefont {Lavault}}, \bibinfo {author} {\bibfnamefont
  {T.}~\bibnamefont {Mairs}}, \bibinfo {author} {\bibfnamefont
  {S.}~\bibnamefont {McSweeney}}, \bibinfo {author} {\bibfnamefont
  {S.}~\bibnamefont {Petitdemange}}, \bibinfo {author} {\bibfnamefont
  {V.}~\bibnamefont {Rey-Bakaikoa}}, \bibinfo {author} {\bibfnamefont
  {J.}~\bibnamefont {Surr}}, \bibinfo {author} {\bibfnamefont {P.}~\bibnamefont
  {Theveneau}}, \bibinfo {author} {\bibfnamefont {G.}~\bibnamefont {Leonard}},
  \ and\ \bibinfo {author} {\bibfnamefont {C.}~\bibnamefont
  {Mueller-Dieckmann}},\ }\bibfield  {title} {{{ID}29: a high-intensity highly
  automated {ESRF} beamline for macromolecular crystallography experiments
  exploiting anomalous scattering},\ }\href {\doibase
  10.1107/S0909049512009715} {\bibfield  {journal} {\bibinfo  {journal} {\emph
  {J. Synchrotron Radiat.}}\ }\textbf {\bibinfo {volume} {19}},\ \bibinfo
  {pages} {455} (\bibinfo {year} {2012})}\BibitemShut {NoStop}%
\bibitem [{\citenamefont {Kraft}\ \emph {et~al.}(2009)\citenamefont {Kraft},
  \citenamefont {Bergamaschi}, \citenamefont {Broennimann}, \citenamefont
  {Dinapoli}, \citenamefont {Eikenberry}, \citenamefont {Henrich},
  \citenamefont {Johnson}, \citenamefont {Mozzanica}, \citenamefont
  {Schlep{\"{u}}tz}, \citenamefont {Willmott},\ and\ \citenamefont
  {Schmitt}}]{kraft_jsr_2009}%
  \BibitemOpen
  \bibfield  {author} {\bibinfo {author} {\bibfnamefont {P.}~\bibnamefont
  {Kraft}}, \bibinfo {author} {\bibfnamefont {A.}~\bibnamefont {Bergamaschi}},
  \bibinfo {author} {\bibfnamefont {C.}~\bibnamefont {Broennimann}}, \bibinfo
  {author} {\bibfnamefont {R.}~\bibnamefont {Dinapoli}}, \bibinfo {author}
  {\bibfnamefont {E.~F.}\ \bibnamefont {Eikenberry}}, \bibinfo {author}
  {\bibfnamefont {B.}~\bibnamefont {Henrich}}, \bibinfo {author} {\bibfnamefont
  {I.}~\bibnamefont {Johnson}}, \bibinfo {author} {\bibfnamefont
  {A.}~\bibnamefont {Mozzanica}}, \bibinfo {author} {\bibfnamefont {C.~M.}\
  \bibnamefont {Schlep{\"{u}}tz}}, \bibinfo {author} {\bibfnamefont {P.~R.}\
  \bibnamefont {Willmott}}, \ and\ \bibinfo {author} {\bibfnamefont
  {B.}~\bibnamefont {Schmitt}},\ }\bibfield  {title} {Performance of
  single-photon-counting {PILATUS} detector modules,\ }\href {\doibase
  10.1107/S0909049509009911} {\bibfield  {journal} {\bibinfo  {journal} {\emph
  {J. Synchrotron Radiat.}}\ }\textbf {\bibinfo {volume} {16}},\ \bibinfo
  {pages} {368} (\bibinfo {year} {2009})}\BibitemShut {NoStop}%
\bibitem [{\citenamefont {Krisch}\ and\ \citenamefont
  {Sette}(2007)}]{kirsch_Springer_2007}%
  \BibitemOpen
  \bibfield  {author} {\bibinfo {author} {\bibfnamefont {M.}~\bibnamefont
  {Krisch}}\ and\ \bibinfo {author} {\bibfnamefont {F.}~\bibnamefont {Sette}},\
  }\href@noop {} {\emph {\bibinfo {title} {Inelastic X-ray Scattering from
  Phonons. Light Scattering in solids, Novel Materials and Techniques, Topics
  in Applied Physics 108}}}\ (\bibinfo  {publisher} {Springer-Verlag},\
  \bibinfo {year} {2007})\BibitemShut {NoStop}%
\bibitem [{\citenamefont {Bosak}\ and\ \citenamefont
  {Krisch}(2005)}]{bosak_prb_2005}%
  \BibitemOpen
  \bibfield  {author} {\bibinfo {author} {\bibfnamefont {A.}~\bibnamefont
  {Bosak}}\ and\ \bibinfo {author} {\bibfnamefont {M.}~\bibnamefont {Krisch}},\
  }\bibfield  {title} {Phonon density of states probed by inelastic x-ray
  scattering,\ }\href {\doibase 10.1103/PhysRevB.72.224305} {\bibfield
  {journal} {\bibinfo  {journal} {\emph {Phys. Rev. B}}\ }\textbf {\bibinfo
  {volume} {72}},\ \bibinfo {pages} {224305} (\bibinfo {year}
  {2005})}\BibitemShut {NoStop}%
\bibitem [{\citenamefont {Gonze}\ and\ \citenamefont
  {Lee}(1997)}]{gonze_prb_1997_2}%
  \BibitemOpen
  \bibfield  {author} {\bibinfo {author} {\bibfnamefont {X.}~\bibnamefont
  {Gonze}}\ and\ \bibinfo {author} {\bibfnamefont {C.}~\bibnamefont {Lee}},\
  }\bibfield  {title} {Dynamical matrices, born effective charges, dielectric
  permittivity tensors, and interatomic force constants from density-functional
  perturbation theory,\ }\href {\doibase 10.1103/PhysRevB.55.10355} {\bibfield
  {journal} {\bibinfo  {journal} {\emph {Phys. Rev. B}}\ }\textbf {\bibinfo
  {volume} {55}},\ \bibinfo {pages} {10355} (\bibinfo {year}
  {1997})}\BibitemShut {NoStop}%
\bibitem [{\citenamefont {Clark}\ \emph {et~al.}(2005)\citenamefont {Clark},
  \citenamefont {Segall}, \citenamefont {Pickard}, \citenamefont {Hasnip},
  \citenamefont {Probert}, \citenamefont {Refson},\ and\ \citenamefont
  {Payne}}]{clark_zkri_2005}%
  \BibitemOpen
  \bibfield  {author} {\bibinfo {author} {\bibfnamefont {S.}~\bibnamefont
  {Clark}}, \bibinfo {author} {\bibfnamefont {M.}~\bibnamefont {Segall}},
  \bibinfo {author} {\bibfnamefont {C.}~\bibnamefont {Pickard}}, \bibinfo
  {author} {\bibfnamefont {P.}~\bibnamefont {Hasnip}}, \bibinfo {author}
  {\bibfnamefont {M.}~\bibnamefont {Probert}}, \bibinfo {author} {\bibfnamefont
  {K.}~\bibnamefont {Refson}}, \ and\ \bibinfo {author} {\bibfnamefont
  {M.}~\bibnamefont {Payne}},\ }\bibfield  {title} {First principles methods
  using {CASTEP},\ }\href {\doibase 10.1524/zkri.220.5.567.65075} {\bibfield
  {journal} {\bibinfo  {journal} {\emph {Z. Kristallogr.}}\ }\textbf {\bibinfo
  {volume} {220}},\ \bibinfo {pages} {567} (\bibinfo {year}
  {2005})}\BibitemShut {NoStop}%
\bibitem [{\citenamefont {Refson}\ \emph {et~al.}(2006)\citenamefont {Refson},
  \citenamefont {Tulip},\ and\ \citenamefont {Clark}}]{refson_prb_2006}%
  \BibitemOpen
  \bibfield  {author} {\bibinfo {author} {\bibfnamefont {K.}~\bibnamefont
  {Refson}}, \bibinfo {author} {\bibfnamefont {P.~R.}\ \bibnamefont {Tulip}}, \
  and\ \bibinfo {author} {\bibfnamefont {S.~J.}\ \bibnamefont {Clark}},\
  }\bibfield  {title} {Variational density-functional perturbation theory for
  dielectrics and lattice dynamics,\ }\href {\doibase
  10.1103/PhysRevB.73.155114} {\bibfield  {journal} {\bibinfo  {journal} {\emph
  {Phys. Rev. B}}\ }\textbf {\bibinfo {volume} {73}},\ \bibinfo {pages}
  {155114} (\bibinfo {year} {2006})}\BibitemShut {NoStop}%
\bibitem [{\citenamefont {Perdew}\ and\ \citenamefont
  {Zunger}(1981)}]{perdew_prb_1981}%
  \BibitemOpen
  \bibfield  {author} {\bibinfo {author} {\bibfnamefont {J.~P.}\ \bibnamefont
  {Perdew}}\ and\ \bibinfo {author} {\bibfnamefont {A.}~\bibnamefont
  {Zunger}},\ }\bibfield  {title} {Self-interaction correction to
  density-functional approximations for many-electron systems,\ }\href
  {\doibase 10.1103/PhysRevB.23.5048} {\bibfield  {journal} {\bibinfo
  {journal} {\emph {Phys. Rev. B}}\ }\textbf {\bibinfo {volume} {23}},\
  \bibinfo {pages} {5048} (\bibinfo {year} {1981})}\BibitemShut {NoStop}%
\bibitem [{\citenamefont {Rappe}\ \emph {et~al.}(1990)\citenamefont {Rappe},
  \citenamefont {Rabe}, \citenamefont {Kaxiras},\ and\ \citenamefont
  {Joannopoulos}}]{rappe_prb_1990}%
  \BibitemOpen
  \bibfield  {author} {\bibinfo {author} {\bibfnamefont {A.~M.}\ \bibnamefont
  {Rappe}}, \bibinfo {author} {\bibfnamefont {K.~M.}\ \bibnamefont {Rabe}},
  \bibinfo {author} {\bibfnamefont {E.}~\bibnamefont {Kaxiras}}, \ and\
  \bibinfo {author} {\bibfnamefont {J.~D.}\ \bibnamefont {Joannopoulos}},\
  }\bibfield  {title} {Optimized pseudopotentials,\ }\href {\doibase
  10.1103/PhysRevB.41.1227} {\bibfield  {journal} {\bibinfo  {journal} {\emph
  {Phys. Rev. B}}\ }\textbf {\bibinfo {volume} {41}},\ \bibinfo {pages} {1227}
  (\bibinfo {year} {1990})}\BibitemShut {NoStop}%
\bibitem [{\citenamefont {Pfrommer}\ \emph {et~al.}(1997)\citenamefont
  {Pfrommer}, \citenamefont {C{\^o}t{\'e}}, \citenamefont {Louie},\ and\
  \citenamefont {Cohen}}]{pfrommer_jcp_1997}%
  \BibitemOpen
  \bibfield  {author} {\bibinfo {author} {\bibfnamefont {B.~G.}\ \bibnamefont
  {Pfrommer}}, \bibinfo {author} {\bibfnamefont {M.}~\bibnamefont
  {C{\^o}t{\'e}}}, \bibinfo {author} {\bibfnamefont {S.~G.}\ \bibnamefont
  {Louie}}, \ and\ \bibinfo {author} {\bibfnamefont {M.~L.}\ \bibnamefont
  {Cohen}},\ }\bibfield  {title} {Relaxation of crystals with the quasi-newton
  method,\ }\href {\doibase 10.1006/jcph.1996.5612} {\bibfield  {journal}
  {\bibinfo  {journal} {\emph {J. Comput. Phys.}}\ }\textbf {\bibinfo {volume}
  {131}},\ \bibinfo {pages} {233 } (\bibinfo {year} {1997})}\BibitemShut
  {NoStop}%
\bibitem [{\citenamefont {Gonze}\ \emph {et~al.}(1994)\citenamefont {Gonze},
  \citenamefont {Charlier}, \citenamefont {Allan},\ and\ \citenamefont
  {Teter}}]{gonze_prb_1994}%
  \BibitemOpen
  \bibfield  {author} {\bibinfo {author} {\bibfnamefont {X.}~\bibnamefont
  {Gonze}}, \bibinfo {author} {\bibfnamefont {J.-C.}\ \bibnamefont {Charlier}},
  \bibinfo {author} {\bibfnamefont {D.}~\bibnamefont {Allan}}, \ and\ \bibinfo
  {author} {\bibfnamefont {M.}~\bibnamefont {Teter}},\ }\bibfield  {title}
  {Interatomic force constants from first principles: The case of
  $\alpha${}-quartz,\ }\href {\doibase 10.1103/PhysRevB.50.13035} {\bibfield
  {journal} {\bibinfo  {journal} {\emph {Phys. Rev. B}}\ }\textbf {\bibinfo
  {volume} {50}},\ \bibinfo {pages} {13035} (\bibinfo {year}
  {1994})}\BibitemShut {NoStop}%
\bibitem [{\citenamefont {Parlinski}\ \emph {et~al.}(1997)\citenamefont
  {Parlinski}, \citenamefont {Li},\ and\ \citenamefont
  {Kawazoe}}]{parlinski_prl_1997}%
  \BibitemOpen
  \bibfield  {author} {\bibinfo {author} {\bibfnamefont {K.}~\bibnamefont
  {Parlinski}}, \bibinfo {author} {\bibfnamefont {Z.~Q.}\ \bibnamefont {Li}}, \
  and\ \bibinfo {author} {\bibfnamefont {Y.}~\bibnamefont {Kawazoe}},\
  }\bibfield  {title} {First-principles determination of the soft mode in cubic
  {Z}r{O}$_2$,\ }\href {\doibase 10.1103/PhysRevLett.78.4063} {\bibfield
  {journal} {\bibinfo  {journal} {\emph {Phys. Rev. Lett.}}\ }\textbf {\bibinfo
  {volume} {78}},\ \bibinfo {pages} {4063} (\bibinfo {year}
  {1997})}\BibitemShut {NoStop}%
\bibitem [{\citenamefont {Xu}\ and\ \citenamefont
  {Chiang}(2005)}]{xu_zkri_2005}%
  \BibitemOpen
  \bibfield  {author} {\bibinfo {author} {\bibfnamefont {R.~Q.}\ \bibnamefont
  {Xu}}\ and\ \bibinfo {author} {\bibfnamefont {T.~C.}\ \bibnamefont
  {Chiang}},\ }\bibfield  {title} {Determination of phonon dispersion relations
  by x-ray thermal diffuse scattering,\ }\href {\doibase
  10.1524/zkri.2005.220.12.1009} {\bibfield  {journal} {\bibinfo  {journal}
  {\emph {Z. Kristallogr.}}\ }\textbf {\bibinfo {volume} {220}},\ \bibinfo
  {pages} {1009} (\bibinfo {year} {2005})}\BibitemShut {NoStop}%
\bibitem [{\citenamefont {Cromer}\ and\ \citenamefont
  {Mann}(1968)}]{cromer_ac_1968}%
  \BibitemOpen
  \bibfield  {author} {\bibinfo {author} {\bibfnamefont {D.~T.}\ \bibnamefont
  {Cromer}}\ and\ \bibinfo {author} {\bibfnamefont {J.~B.}\ \bibnamefont
  {Mann}},\ }\bibfield  {title} {{X-ray scattering factors computed from
  numerical {H}artree{--}{F}ock wave functions},\ }\href {\doibase
  10.1107/S0567739468000550} {\bibfield  {journal} {\bibinfo  {journal} {\emph
  {Acta Crystallogr., Sect. A}}\ }\textbf {\bibinfo {volume} {24}},\ \bibinfo
  {pages} {321} (\bibinfo {year} {1968})}\BibitemShut {NoStop}%
\bibitem [{\citenamefont {Bosak}\ \emph {et~al.}(2012)\citenamefont {Bosak},
  \citenamefont {Krisch}, \citenamefont {Chernyshov}, \citenamefont {Winkler},
  \citenamefont {Milman}, \citenamefont {Refson},\ and\ \citenamefont
  {Schulze-Briese}}]{bosak_zkri_2012}%
  \BibitemOpen
  \bibfield  {author} {\bibinfo {author} {\bibfnamefont {A.}~\bibnamefont
  {Bosak}}, \bibinfo {author} {\bibfnamefont {M.}~\bibnamefont {Krisch}},
  \bibinfo {author} {\bibfnamefont {D.}~\bibnamefont {Chernyshov}}, \bibinfo
  {author} {\bibfnamefont {B.}~\bibnamefont {Winkler}}, \bibinfo {author}
  {\bibfnamefont {V.}~\bibnamefont {Milman}}, \bibinfo {author} {\bibfnamefont
  {K.}~\bibnamefont {Refson}}, \ and\ \bibinfo {author} {\bibfnamefont
  {C.}~\bibnamefont {Schulze-Briese}},\ }\bibfield  {title} {New insights into
  the lattice dynamics of $\alpha$-quartz,\ }\href {\doibase
  10.1524/zkri.2012.1432} {\bibfield  {journal} {\bibinfo  {journal} {\emph {Z.
  Kristallogr.}}\ }\textbf {\bibinfo {volume} {227}},\ \bibinfo {pages} {84}
  (\bibinfo {year} {2012})}\BibitemShut {NoStop}%
\bibitem [{\citenamefont {Wehinger}\ \emph {et~al.}(2013)\citenamefont
  {Wehinger}, \citenamefont {Bosak}, \citenamefont {Chumakov}, \citenamefont
  {Mirone}, \citenamefont {Winkler}, \citenamefont {Dubrovinsky}, \citenamefont
  {Dubrovinskaia}, \citenamefont {Brazhkin}, \citenamefont {Dyuzheva},\ and\
  \citenamefont {Krisch}}]{wehinger_jpcm_2013}%
  \BibitemOpen
  \bibfield  {author} {\bibinfo {author} {\bibfnamefont {B.}~\bibnamefont
  {Wehinger}}, \bibinfo {author} {\bibfnamefont {A.}~\bibnamefont {Bosak}},
  \bibinfo {author} {\bibfnamefont {A.}~\bibnamefont {Chumakov}}, \bibinfo
  {author} {\bibfnamefont {A.}~\bibnamefont {Mirone}}, \bibinfo {author}
  {\bibfnamefont {B.}~\bibnamefont {Winkler}}, \bibinfo {author} {\bibfnamefont
  {L.}~\bibnamefont {Dubrovinsky}}, \bibinfo {author} {\bibfnamefont
  {N.}~\bibnamefont {Dubrovinskaia}}, \bibinfo {author} {\bibfnamefont
  {V.}~\bibnamefont {Brazhkin}}, \bibinfo {author} {\bibfnamefont
  {T.}~\bibnamefont {Dyuzheva}}, \ and\ \bibinfo {author} {\bibfnamefont
  {M.}~\bibnamefont {Krisch}},\ }\bibfield  {title} {Lattice dynamics of
  coesite,\ }\href {\doibase 10.1088/0953-8984/25/27/275401} {\bibfield
  {journal} {\bibinfo  {journal} {\emph {J. Phys.: Condens. Matter}}\ }\textbf
  {\bibinfo {volume} {25}},\ \bibinfo {pages} {275401} (\bibinfo {year}
  {2013})}\BibitemShut {NoStop}%
\bibitem [{\citenamefont {Bosak}\ \emph {et~al.}(2009)\citenamefont {Bosak},
  \citenamefont {Fischer}, \citenamefont {Krisch}, \citenamefont {Brazhkin},
  \citenamefont {Dyuzheva}, \citenamefont {Winkler}, \citenamefont {Wilson},
  \citenamefont {Weidner}, \citenamefont {Refson},\ and\ \citenamefont
  {Milman}}]{bosak_grl_2009}%
  \BibitemOpen
  \bibfield  {author} {\bibinfo {author} {\bibfnamefont {A.}~\bibnamefont
  {Bosak}}, \bibinfo {author} {\bibfnamefont {I.}~\bibnamefont {Fischer}},
  \bibinfo {author} {\bibfnamefont {M.}~\bibnamefont {Krisch}}, \bibinfo
  {author} {\bibfnamefont {V.}~\bibnamefont {Brazhkin}}, \bibinfo {author}
  {\bibfnamefont {T.}~\bibnamefont {Dyuzheva}}, \bibinfo {author}
  {\bibfnamefont {B.}~\bibnamefont {Winkler}}, \bibinfo {author} {\bibfnamefont
  {D.}~\bibnamefont {Wilson}}, \bibinfo {author} {\bibfnamefont
  {D.}~\bibnamefont {Weidner}}, \bibinfo {author} {\bibfnamefont
  {K.}~\bibnamefont {Refson}}, \ and\ \bibinfo {author} {\bibfnamefont
  {V.}~\bibnamefont {Milman}},\ }\bibfield  {title} {Lattice dynamics of
  stishovite from powder inelastic x-ray scattering,\ }\href {\doibase
  10.1029/2009GL040257} {\bibfield  {journal} {\bibinfo  {journal} {\emph
  {Geophys. Res. Lett.}}\ }\textbf {\bibinfo {volume} {36}},\ \bibinfo {pages}
  {L19309} (\bibinfo {year} {2009})}\BibitemShut {NoStop}%
\bibitem [{\citenamefont {Van~Hove}(1953)}]{vanhove_pr_1953}%
  \BibitemOpen
  \bibfield  {author} {\bibinfo {author} {\bibfnamefont {L.}~\bibnamefont
  {Van~Hove}},\ }\bibfield  {title} {The occurrence of singularities in the
  elastic frequency distribution of a crystal,\ }\href {\doibase
  10.1103/PhysRev.89.1189} {\bibfield  {journal} {\bibinfo  {journal} {\emph
  {Phys. Rev.}}\ }\textbf {\bibinfo {volume} {89}},\ \bibinfo {pages} {1189}
  (\bibinfo {year} {1953})}\BibitemShut {NoStop}%
\bibitem [{\citenamefont {He}\ \emph {et~al.}(2014)\citenamefont {He},
  \citenamefont {Liu}, \citenamefont {Hautier}, \citenamefont {Oliveira},
  \citenamefont {Marques}, \citenamefont {Vila}, \citenamefont {Rehr},
  \citenamefont {Rignanese},\ and\ \citenamefont {Zhou}}]{he_prb_2014}%
  \BibitemOpen
  \bibfield  {author} {\bibinfo {author} {\bibfnamefont {L.}~\bibnamefont
  {He}}, \bibinfo {author} {\bibfnamefont {F.}~\bibnamefont {Liu}}, \bibinfo
  {author} {\bibfnamefont {G.}~\bibnamefont {Hautier}}, \bibinfo {author}
  {\bibfnamefont {M.~J.~T.}\ \bibnamefont {Oliveira}}, \bibinfo {author}
  {\bibfnamefont {M.~A.~L.}\ \bibnamefont {Marques}}, \bibinfo {author}
  {\bibfnamefont {F.~D.}\ \bibnamefont {Vila}}, \bibinfo {author}
  {\bibfnamefont {J.~J.}\ \bibnamefont {Rehr}}, \bibinfo {author}
  {\bibfnamefont {G.-M.}\ \bibnamefont {Rignanese}}, \ and\ \bibinfo {author}
  {\bibfnamefont {A.}~\bibnamefont {Zhou}},\ }\bibfield  {title} {Accuracy of
  generalized gradient approximation functionals for density-functional
  perturbation theory calculations,\ }\href {\doibase
  10.1103/PhysRevB.89.064305} {\bibfield  {journal} {\bibinfo  {journal} {\emph
  {Phys. Rev. B}}\ }\textbf {\bibinfo {volume} {89}},\ \bibinfo {pages}
  {064305} (\bibinfo {year} {2014})}\BibitemShut {NoStop}%
\bibitem [{\citenamefont {Dubrovinsky}\ \emph {et~al.}(2001)\citenamefont
  {Dubrovinsky}, \citenamefont {Dubrovinskaia}, \citenamefont {Saxena},
  \citenamefont {Tutti}, \citenamefont {Rekhi}, \citenamefont {Bihan},
  \citenamefont {Shen},\ and\ \citenamefont {Hu}}]{dubrovinsky_cpl_2001}%
  \BibitemOpen
  \bibfield  {author} {\bibinfo {author} {\bibfnamefont {L.}~\bibnamefont
  {Dubrovinsky}}, \bibinfo {author} {\bibfnamefont {N.}~\bibnamefont
  {Dubrovinskaia}}, \bibinfo {author} {\bibfnamefont {S.}~\bibnamefont
  {Saxena}}, \bibinfo {author} {\bibfnamefont {F.}~\bibnamefont {Tutti}},
  \bibinfo {author} {\bibfnamefont {S.}~\bibnamefont {Rekhi}}, \bibinfo
  {author} {\bibfnamefont {T.~L.}\ \bibnamefont {Bihan}}, \bibinfo {author}
  {\bibfnamefont {G.}~\bibnamefont {Shen}}, \ and\ \bibinfo {author}
  {\bibfnamefont {J.}~\bibnamefont {Hu}},\ }\bibfield  {title}
  {Pressure-induced transformations of cristobalite,\ }\href {\doibase
  http://dx.doi.org/10.1016/S0009-2614(00)01147-7} {\bibfield  {journal}
  {\bibinfo  {journal} {\emph {Chem. Phys. Lett.}}\ }\textbf {\bibinfo {volume}
  {333}},\ \bibinfo {pages} {264 } (\bibinfo {year} {2001})}\BibitemShut
  {NoStop}%
\bibitem [{\citenamefont {Dubrovinsky}\ \emph {et~al.}(2004)\citenamefont
  {Dubrovinsky}, \citenamefont {Dubrovinskaia}, \citenamefont {Prakapenka},
  \citenamefont {Seifert}, \citenamefont {Langenhorst}, \citenamefont
  {Dmitriev}, \citenamefont {Weber},\ and\ \citenamefont
  {Le~Bihan}}]{dubrovinsky_pepi_2004}%
  \BibitemOpen
  \bibfield  {author} {\bibinfo {author} {\bibfnamefont {L.~S.}\ \bibnamefont
  {Dubrovinsky}}, \bibinfo {author} {\bibfnamefont {N.~A.}\ \bibnamefont
  {Dubrovinskaia}}, \bibinfo {author} {\bibfnamefont {V.}~\bibnamefont
  {Prakapenka}}, \bibinfo {author} {\bibfnamefont {F.}~\bibnamefont {Seifert}},
  \bibinfo {author} {\bibfnamefont {F.}~\bibnamefont {Langenhorst}}, \bibinfo
  {author} {\bibfnamefont {V.}~\bibnamefont {Dmitriev}}, \bibinfo {author}
  {\bibfnamefont {H.~P.}\ \bibnamefont {Weber}}, \ and\ \bibinfo {author}
  {\bibfnamefont {T.}~\bibnamefont {Le~Bihan}},\ }\bibfield  {title} {A class
  of new high-pressure silica polymorphs,\ }\href {\doibase
  10.1016/j.pepi.2003.06.006} {\bibfield  {journal} {\bibinfo  {journal} {\emph
  {Phys. Earth Planet. Inter.}}\ }\textbf {\bibinfo {volume} {143--144}},\
  \bibinfo {pages} {231 } (\bibinfo {year} {2004})}\BibitemShut {NoStop}%
\bibitem [{\citenamefont {Kimizuka}\ \emph {et~al.}(2005)\citenamefont
  {Kimizuka}, \citenamefont {Ogata},\ and\ \citenamefont
  {Shibutani}}]{kimizuka_mt_2005}%
  \BibitemOpen
  \bibfield  {author} {\bibinfo {author} {\bibfnamefont {H.}~\bibnamefont
  {Kimizuka}}, \bibinfo {author} {\bibfnamefont {S.}~\bibnamefont {Ogata}}, \
  and\ \bibinfo {author} {\bibfnamefont {Y.}~\bibnamefont {Shibutani}},\
  }\bibfield  {title} {High-pressure elasticity and auxetic property of
  $\alpha$-cristobalite,\ }\href {\doibase 10.2320/matertrans.46.1161}
  {\bibfield  {journal} {\bibinfo  {journal} {\emph {Mater. Trans.}}\ }\textbf
  {\bibinfo {volume} {46}},\ \bibinfo {pages} {1161} (\bibinfo {year}
  {2005})}\BibitemShut {NoStop}%
\end{thebibliography}%

\end{document}